\documentclass[sigconf, acm]{acmart}

\usepackage{booktabs, makecell, multirow, tabularx}
\usepackage{adjustbox}
\usepackage[dvipsnames]{xcolor}
\usepackage{enumitem}
\usepackage[skins,breakable]{tcolorbox}
\usepackage[table]{xcolor}
\usepackage{subcaption}
\usepackage{xfrac}
\usepackage{algorithm}
\usepackage{algpseudocode}
\usepackage{balance}
\usepackage{soul}

\newtheorem{theorem}{Theorem}
\newtheorem{lemma}{Lemma}
\newtheorem{proposition}{Proposition}
\newtheorem{assumption}{Assumption}
\theoremstyle{definition}

\newcommand{\exampleblock}[2]{
    \par \vspace{0.1em} 
    \noindent\textbf{Example #1:} \textit{#2}
    \vspace{0.1em} 
}

\newcommand{\spara}[1]{\smallskip\noindent{\bf #1}}

\setcopyright{cc}
\setcctype{by}
\copyrightyear{} 
\acmConference[SIGMOD '27]{}{June 13--19, 2027}{Huntington Beach, CA, USA}
\acmISBN{}
\acmDOI{}

\title{Cost-Efficient RAG for Entity Matching with LLMs: A~Blocking-based Batch Processing Framework}

\author{Chuangtao Ma}
\authornote{Equal contribution.}
\affiliation{%
  \institution{Aalborg University}
  \city{Aalborg}
  \country{Denmark}
}
\email{chuma@cs.aau.dk}

\author{Zeyu Zhang}
\authornotemark[1]
\affiliation{%
  \institution{University of Amsterdam}
  \city{Amsterdam}
  \country{The Netherlands}
}
\email{z.zhang2@uva.nl}

\author{Arijit Khan}
\affiliation{%
  \institution{Bowling Green State University, USA} \institution{Aalborg University, Denmark}
  \country{}
}
\email{arijitk@bgsu.edu}

\author{Sebastian Schelter}
\affiliation{%
  \institution{BIFOLD \& TU Berlin}
  \city{Berlin}
  \country{Germany}
}
\email{schelter@tu-berlin.de}

\author{Paul Groth}
\affiliation{%
  \institution{University of Amsterdam}
  \city{Amsterdam}
  \country{The Netherlands}
}
\email{p.t.groth@uva.nl}

\begin{document}


\begin{abstract} 
Large language models (LLMs) offer a compelling approach to entity matching (EM) in label-scarce settings, yet their matching quality is sensitive to records with ambiguous or sparse attributes due to the absence of external factual grounding. 
Retrieval-augmented generation (RAG) directly addresses this grounding gap by retrieving and injecting relevant external knowledge.
However, naively applying RAG to large-scale EM is expensive due to per-pair retrieval calls over large knowledge graphs (KGs) and the increase in input tokens caused by enriching prompts with retrieved contextual knowledge. 
To address the above limitations, we introduce the \textit{CE-RAG4EM} (\underline{C}ost-\underline{E}fficient \underline{RAG} for \underline{E}ntity \underline{M}atching) framework that reduces overhead through blocking-based batch retrieval and generation.
We investigate the benefits of this framework through a systematic analysis of RAG systems for entity matching, focusing on blocking-aware batch optimizations and retrieval granularity.  
Extensive experiments on nine widely used real-world entity matching benchmark datasets show that \textit{CE-RAG4EM} under a zero-shot setting achieves comparable or improved matching quality while substantially reducing end-to-end overhead. 
The results demonstrate that our \textit{CE-RAG4EM} closes the knowledge grounding gap of the LLM-based entity matching (LLM-EM) baseline and yields significant improvements in F1 scores on datasets with ambiguous or numerically sparse attributes, while its batch retrieval and inference substantially reduce the overhead and maintain competitive matching quality compared to na\"{i}ve RAG4EM.
Our analysis identifies clear performance--overhead trade‑offs across blocking‑based batch sizes and retrieval granularity, showing how matching quality and computational cost shift with each design choice. These findings offer guidance for building scalable, efficient RAG systems for entity matching that balance accuracy and cost.
\end{abstract}

\maketitle

\section{Introduction}
\label{sec:intro}
Entity matching (EM) is a fundamental data integration task that determines whether two records refer to the same real-world entity~\cite{PanahiWDN17, SinghMEMPQST17, MudgalLRDPKDAR18}. 
Large language models (LLMs) such as GPT-4 have emerged as a compelling approach to EM, offering strong generalization without requiring task-specific labeled training data~\cite{PeetersSB25, ZhangGCS25, WangCLCHSWZ25}. 
This is a significant practical advantage over solutions based on pre-trained language models (PLMs), e.g., Ditto~\cite{LiDeepEM20, LiLSDT23} and Unicorn~\cite{TuFTWL0JG23}, which achieve strong results but require fine-tuning on labeled entity pairs that are often unavailable or expensive to annotate in real-world enterprise data engineering. 
However, deploying LLMs for real-world entity matching remains challenging. Prior work shows that matching quality degrades on large, heterogeneous tables with mixed attribute types, missing values, and highly imbalanced match distributions~\cite{BodensohnUnveil2025, PeetersB23, chen2026generalized}. LLM-based methods are also sensitive to prompt design and vulnerable to hallucination and model bias~\cite{PeetersSB25}, limiting their robustness in practical deployment.

Beyond these known limitations, we observe a characteristic precision--recall imbalance in LLM-based EM. When prompted to provide a reasoning path, LLMs maintain relatively stable precision but exhibit consistently lower recall on datasets containing ambiguous, sparse, or externally grounded records. In such cases, the model tends to reject matches when the input records alone provide insufficient evidence, resulting in false negatives. 
We interpret this behavior as an EM-specific \emph{grounding gap}~\cite{Harnad2023LLM, ShaikhGKGYJ2023}: the information contained in the input records is insufficient for the LLM to confidently establish an otherwise valid match. This perspective unifies previously reported issues such as ambiguity, sparse attributes, and false negatives under insufficient factual grounding. As shown in \S\ref{sec:experiments}, this failure pattern is consistent across datasets and LLMs.

Retrieval-augmented generation (RAG)~\cite{LewisPPPKGKLYR020, KLZZZ26} offers a direct way to mitigate the grounding gap by injecting relevant external knowledge at inference time. Rather than relying solely on the LLM's parametric knowledge, RAG for Entity Matching (RAG4EM) augments each record pair with facts retrieved from a knowledge base, such as Wikidata, providing additional evidence to resolve ambiguous matches.
As illustrated below, retrieved evidence chains from Wikidata can supply the factual grounding needed for LLMs to support a correct matching decision.
\exampleblock{}{
\noindent Consider the record pair \texttt{``yujean kang's gourmet chinese cuisine''} (city: \textit{Los Angeles}) (type: \textit{asian}) and \texttt{``yujean kang's''} (city: \textit{Pasadena}) (type: \textit{chinese}). Given only the input records, the LLM incorrectly predicts a non-match, despite semantically equivalent addresses and phone numbers, because of the apparent discrepancies in city and restaurant type. RAG4EM retrieves external evidence such as ``Pasadena $\rightarrow$ \textsf{instance of} $\rightarrow$ city of California $\rightarrow$ \textsf{located in} $\rightarrow$ Los Angeles County'' and ``Chinese restaurant $\rightarrow$ \textsf{subclass of} $\rightarrow$ Asian restaurant.'' These evidence chains provide geographic and semantic context for reconciling the conflicting attribute values, helping the LLM correctly identify the pair as a match.
}

\begin{figure}[htbp]
  \centering
  \begin{subfigure}[t]{1\linewidth}
    \centering
    \includegraphics[width=\linewidth]{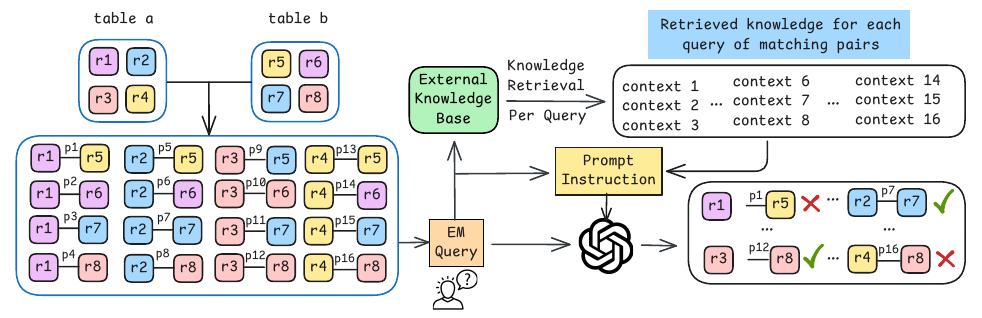}
    \caption{Vanilla RAG4EM per-query retrieval/generation.}
    \label{fig:vanlila-rag4em}
  \end{subfigure}

  \begin{subfigure}[t]{1\linewidth}
    \centering
    \includegraphics[width=\linewidth]{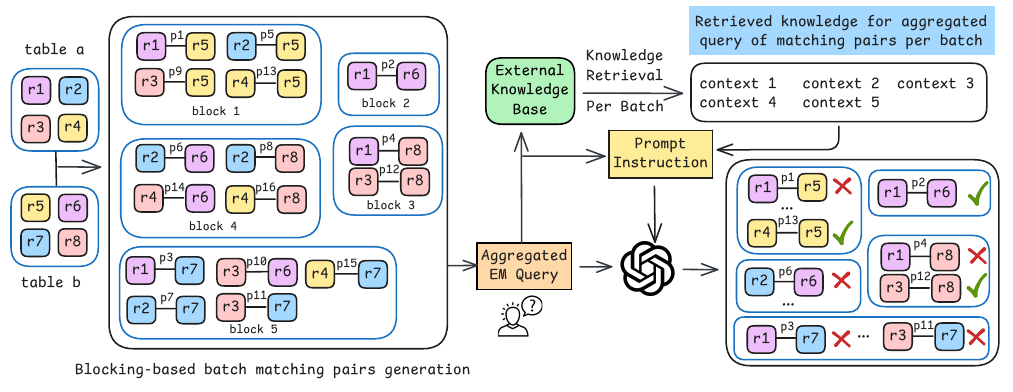}
    \caption{\textit{CE-RAG4EM} with blocking-based batch retrieval/generation.}
    \label{fig:ce-rag4em}
  \end{subfigure}
  \caption{Vanilla RAG4EM vs.\ \textit{CE-RAG4EM}. Matched records share the same color.}
  \label{fig:comparison}
\end{figure}

However, {\em RAG4EM remains largely unexplored}, primarily because applying vanilla RAG to large-scale EM tasks is prohibitively expensive. 
Specifically, comparing all \(m\) records in one table with \(n\) records in another incurs a quadratic \(O(mn)\) cost. 
As shown in Figure~\ref{fig:vanlila-rag4em}, vanilla RAG4EM naively compares 4 records in table a to 4 records in table b, requiring 16 distinct comparisons, where each comparison incurs independent retrieval and LLM inference costs. 

There are two major sources of overhead introduced when directly applying vanilla RAG4EM.
\textbf{(D1) Retrieval overhead:} Vanilla RAG issues one retrieval call per candidate pair, resulting in $N$ retrieval calls, where $N$ is the number of candidate pairs.
Over large knowledge bases, each call involves query encoding and approximate nearest-neighbor search, introducing a non-trivial per-call cost that accumulates as $N$ grows.
\textbf{(D2) Inference overhead:} The retrieved context increases the length of each LLM prompt. Each of the $N$ pairs incurs the cost of processing the pair, the task instruction, and the retrieved context.
Both sources of overhead scale linearly with $N$, making per-pair RAG4EM impractical for large-scale EM.
Moreover, neither standard RAG nor its graph‑augmented variants (e.g., GraphRAG \cite{HuLZPLZ25}, KG‑RAG \cite{ZhuXLLH25}) have been systematically studied in LLM‑based EM, leaving their retrieval and graph traversal strategies largely unexplored for EM.

To address these challenges, we introduce the \textit{CE-RAG4EM} (\underline{C}ost-\underline{E}fficient \underline{RAG} for \underline{E}ntity \underline{M}atching) framework for zero-shot entity matching that amortizes retrieval and generation costs through blocking-based batching. Figure~\ref{fig:comparison} compares vanilla RAG4EM with our proposed \textit{CE-RAG4EM}. Vanilla RAG4EM performs retrieval and generation 
independently for each candidate pair, whereas \textit{CE-RAG4EM} performs these operations at the batch level. We consider a zero-shot setting in which no task-specific labeled data or in-context demonstrations are available, and the LLM directly performs matching across two structured data sources.

While prior work has explored batch generation to reduce LLM inference cost~\cite{fan2024cost, fu2025context}, \textit{CE-RAG4EM} does not treat batching as a standalone inference optimization. Instead, it reuses the intermediate groups naturally produced by the EM blocking stage to construct semantically coherent batches for both retrieval and generation.
Traditional entity matching methods~\cite{PKPN14, LiDeepEM20, TuFTWL0JG23} 
primarily use blocking to prune unlikely matches and reduce the comparison space.  
In \textit{CE-RAG4EM}, we 
additionally use blocking-derived groups as semantically coherent batching units. These groups organize related candidate pairs so that retrieval and generation can both be amortized across multiple pairs while preserving a coherent shared context.
Specifically, a single aggregated retrieval operation serves all pairs within a batch, avoiding $N$ independent per-pair retrieval calls and thereby addressing D1. Similarly, a single batch-level prompt call jointly processes all pairs using the shared retrieved context, amortizing the instruction and context overhead across the batch and thereby addressing D2.

Concretely, for input tokens, per-pair RAG4EM incurs a cost of
$N \cdot (T_{\text{template}} + T_{\text{context}} + T_{\text{pair}})$,
whereas \textit{CE-RAG4EM} reduces this to
$\left\lceil N/\bar{b} \right\rceil \cdot (T_{\text{template}} + T_{\text{context}}) + N \cdot T_{\text{pair}}$, where $\bar{b}$ denotes the average batch size across all batches, yielding a factor-$\bar{b}$ reduction in the dominant shared overhead of the template and retrieved context. For output tokens, an analogous closed-form reduction is not guaranteed because the output length depends on the LLM's generation behavior. Nevertheless, as shown empirically in \S\ref{sec:experiments}, batch generation in \textit{CE-RAG4EM} also reduces output-token cost.

We further characterize the resulting effectiveness--efficiency trade-off by analyzing how batch size affects evidence coverage, retrieval noise, matching quality, and computational cost.
 
\noindent \textbf{Contributions.} Our main contributions are summarized below.
\begin{itemize}
    \item \textbf{Grounding gap characterization and RAG4EM.} 
    We identify and characterize an EM-specific \emph{grounding gap}, where insufficient factual context leads LLM-based matchers to miss valid matches, particularly for ambiguous and sparse records.
    We introduce \textit{RAG4EM}, which augments record pairs with external knowledge to mitigate this failure mode (\S\ref{sec:rag4em}, \S\ref{subsec:A}).
    
    \item \textbf{CE-RAG4EM with blocking-aware batching.} We introduce \textit{CE-RAG4EM}, which can reuse groups produced by EM blocking as coherent batches for both retrieval and generation. This design amortizes retrieval and LLM inference costs across multiple candidate pairs without introducing a separate batching stage. We further characterize the resulting effectiveness--efficiency trade-off by analyzing how batch size affects retrieval cost, evidence coverage, retrieval noise, and matching quality (\S\ref{subsec:blocking_pair_generation}--\S\ref{subsec:theory}).

    \item \textbf{Comprehensive empirical validation.} 
    We evaluate \textit{CE-RAG4EM} across 9 real-world datasets, 4 LLM backbones, and 3 blocking methods. The results show that external knowledge reliably improves matching effectiveness (\S\ref{subsec:A}), and that blocking-aware batching substantially reduces retrieval and inference cost with only a modest change in matching quality (\S\ref{subsec:B}). We also study the impact of retrieval granularity and demonstrate blocking-induced effectiveness–efficiency trade-offs (\S\ref{subsec:C}, \S\ref{subsec:D}).

\end{itemize}

\section{Preliminaries}
\newcommand{\Retr}{\mathsf{Retr}}
\newcommand{\Serializer}{\mathsf{Serializer}}
\newtheorem{remark}{Remark}
\newtheorem{defn}{Definition}

We introduce the notation and background for LLM-based entity matching (EM), retrieval-augmented generation (RAG), and knowledge graph (KG)-based RAG for EM.

\subsection{LLM-based Entity Matching}
\label{subsec:llm-em}

\begin{defn}[Entity Matching]
\label{def:em}
Let \(T_s\) and \(T_t\) denote a source and a target table, respectively, and let \(\mathcal{R}\) denote the record space.
Given two records \(r_1 \in T_s\) and \(r_2 \in T_t\), entity matching determines whether $r_1$ and $r_2$ refer to the same real-world entity.
\end{defn}
A conventional EM system learns a function \(f: T_s \times T_t \rightarrow \{0,1\}\), where \(f(r_1,r_2)=1\) indicates a match and \(f(r_1,r_2)=0\) otherwise.

\begin{defn}[LLM-based Entity Matching]
\label{def:llm-em}
LLM-based EM replaces the task-specific classifier with a generative model. Given a serialized representation \(x(r_1,r_2)\) of a record pair, the LLM produces a binary matching decision: 
\(
\mathcal{F}_{\mathrm{LLM}}\bigl(x(r_1,r_2)\bigr)
\in \{\text{yes},\text{no}\}.
\)
\end{defn}
Instead of learning a task-specific decision boundary, the LLM is prompted to generate a binary response, namely, yes or no.

\subsection{RAG and KG-RAG for EM} 
\label{subsec:rag_kg-rag-em}

Beyond the records, we assume access to an external factual KG (e.g., Wikidata~\cite{WikidataVrandecicPK23}), containing entity-centric facts and relationships.

\begin{defn}[Knowledge Graph]
\label{def:kg}
A knowledge graph is denoted as
\(\mathcal{G} = (V, P, E)\), where
\(V\) is a finite set of nodes~(entities or concepts),
\(P\) is a finite set of predicate~(relation) types, and \(E \subseteq V \times P \times V\) is a finite set of directed, typed edges represented as triples.
 \end{defn}
Given a record pair \((r_1,r_2)\), RAG-based EM retrieves contextual knowledge from \(\mathcal{G}\) to complement the information contained in the records. Depending on the retrieval granularity, this context may consist of relevant entities, predicates, or KG triples.

\begin{defn}[Knowledge Retriever]
\label{def:knolwedge-retriever}
A knowledge retriever is a function \(\Retr:\mathcal{R}\times\mathcal{R}\rightarrow 2^{\mathcal{K}},\) where \(\mathcal{K}\) denotes the space of retrievable knowledge units. For a record pair \((r_1, r_2)\), \(\Retr(r_1,r_2)\) returns a finite set of knowledge units relevant to the pair.
\end{defn}
The knowledge units in \(\mathcal{K}\) may be represented in different granularities: (1) entity-level context associated with \(r_1\) or \(r_2\), (2) predicate-level context describing relevant relations, or (3) triple-level context obtained by expanding relevant entities or predicates in \(\mathcal{G}\).

\begin{defn}[Graph-Aware Serializer]
\label{def:graph-serializer}
Let \(\Sigma\) be a finite alphabet and \(\Sigma^{*}\) denote the set of all finite sequences over \(\Sigma\). A graph-aware serializer maps a record pair and its retrieved context to a textual sequence: \(x(r_1,r_2) \;=\;\Serializer\bigl(r_1,r_2,\Retr(r_1,r_2)\bigr) \in \Sigma^{*}.\)
\end{defn}
The serializer is responsible for constructing a prompt or input string that exposes
(1) the original attributes of $r_1$ and $r_2$, and
(2) the extracted knowledge $\Retr(r_1,r_2)$,
to the matching model in a structured way (e.g., by listing the record attributes followed by a formatted list of relevant triples).

\begin{defn}[RAG and KG-RAG for EM]
\label{def:rag4em_kg-rag4em}
RAG4EM augments LLM-based EM with retrieved external context. In RAG4EM, the retrieved context is represented as unstructured entity- or predicate-level text, whereas in \textit{KG-RAG4EM} it is formatted explicitly as KG triples.
\end{defn}
Both variants allow the LLM to leverage information beyond the input records, while KG-RAG4EM additionally exposes explicit relational structure through retrieved triples.

\subsection{RAG4EM with Batch Generation}
\label{subsec:rag-batch-input-generation}

We next formalize batch generation, in which the LLM processes multiple candidate record pairs in a single request. 

\begin{defn}[RAG4EM with Batch Generation]
\label{def:rag4em-batch}
Let \(
\mathcal{B} = \bigl(({r_1}^{(1)}, {r_2}^{(1)}), \dots, (r_1^{(b)},r_2^{(b)})\bigr)
\in (\mathcal{R} \times \mathcal{R})^b,
\)
denote a batch of \(b\) record pairs. We construct a single serialized input
\[
bx = \Serializer\bigl(\{(r_1^{(i)},r_2^{(i)})\}_{i=1}^b,\ \{\Retr(r_1^{(i)},r_2^{(i)})\}_{i=1}^b\bigr).
\]
A generative matching model $M$ takes $bx$ as input and produces an output text, which is parsed into a batch of match decisions at once.

\[
\mathcal{A} = \mathcal{F}_{\text{LLM}}(bx)
= (a_1, a_2, \dots, a_b),
\quad a_i \in \{\text{yes}, \text{no}\}.
\]
\end{defn}

Thus, batch generation amortizes a single LLM invocation across multiple candidate pairs, preserving a one-to-one correspondence between input pairs and output decisions.

\noindent\textbf{Remark.} The matching LLM already internalizes substantial knowledge that may overlap with the knowledge encoded in $\mathcal{G}$. Consequently, the extracted  $\Retr(r_1,r_2)$ from a KG can be \emph{beneficial} when it provides factual, precise, or more up-to-date knowledge than the model's internal knowledge; or \emph{redundant} (or even detrimental), when it merely repeats what the model already ``knows'' or introduces noise.
Our formalization allows both: when $\Retr(r_1,r_2) = \emptyset$, or when $M$ effectively ignores the triples in $\Retr(r_1,r_2)$, the system reduces to an LLM-based EM without external knowledge.

\raggedbottom

\section{From Knowledge Grounding to CE-RAG4EM}
\label{sec:rag4em}
This section motivates knowledge grounding for LLM-based entity matching. We first characterize the grounding gap that arises when
the information contained in the input records and the LLM’s parametric knowledge is insufficient for reliable matching. We then present a generic knowledge-grounded EM formulation based on external knowledge retrieval. This formulation motivates the design of \textit{CE-RAG4EM}, a cost-efficient RAG for entity matching, which will be presented in \S~\ref{sec:methodology}.

\subsection{Motivation: The Grounding Gap}
\label{subsec:grounding_gap}

LLM-based EM reasons over serialized record pairs using the information in the input records together with the model's parametric knowledge. This works well when records provide sufficient evidence, but becomes challenging for ambiguous records, sparse attributes, or entity names requiring additional disambiguation. In such cases, the model lacks the factual grounding needed to confidently identify true matches and tends to default to conservative non-match decisions, leading to false negatives. Such a failure mode is often referred to as the \emph{grounding gap}~\cite{Harnad2023LLM, ShaikhGKGYJ2023}.

The severity of the grounding gap depends on the dataset rather than being uniform across LLM-based EM tasks. As we show in \S\ref{sec:experiments} (Table~\ref{tab:cerag4em_llmem_metrics}), on datasets with ambiguous or numerically sparse attributes (e.g., DBGO and ITAM), LLM-based EM attains only 49--50\% recall despite precision above 90\%. In contrast, on well-structured datasets with distinctive attributes, recall is substantially higher. This pattern is consistent across four LLMs and motivates augmenting LLM-EM with external knowledge at inference time.

\subsection{Knowledge-Grounded EM}
\label{subsec:rag4em_pipeline}

A natural way to address the grounding gap is to augment each EM query with context retrieved from an external knowledge base $\mathcal{G}$, such as Wikidata. At a high level, such a knowledge-grounded EM pipeline consists of three stages.

\noindent\textbf{Retrieval.}
Given a candidate pair \((r_1, r_2)\) and the serialized query, dense retrieval is used to obtain the Top-$k$ relevant entities $V_k$ and predicates $P_k$ from $\mathcal{G}$ via approximate nearest-neighbor search (Definition~\ref{def:knolwedge-retriever}). Retrieved entities or predicates can further serve as seeds for graph traversal, such as BFS or neighborhood expansion, to construct structured triples.

\noindent\textbf{Enrichment and Refinement.}
Retrieved symbolic entities, predicates, or triples are enriched with textual descriptions when available and refined using similarity-based ranking and instruction-guided selection to retain context relevant to the candidate pair.

\noindent\textbf{Knowledge-Augmented Inference.}
The refined context is serialized together with the record pair using the graph-aware serializer (Definition~\ref{def:graph-serializer}). The resulting sequence \(x(r_1,r_2)\) is then processed by the LLM to produce a matching decision \(\mathcal{F}_{\text{LLM}}(x(r_1,r_2))
\in \{\text{yes},\text{no}\}.\)
This formulation improves factual grounding, but applying retrieval and generation independently to every candidate pair introduces substantial overhead.

\subsection{Cost of Per-Query Knowledge Grounding}
\label{subsec:rag4em_cost}

A straightforward knowledge-grounded EM pipeline performs retrieval and generation for every candidate pair. For $N$ candidate pairs, this introduces two major sources of overhead:

\begin{itemize}[leftmargin=*]
    \item \textbf{(D1) Retrieval cost.} It issues $N$ independent retrieval operations.
    \item \textbf{(D2) Inference cost.} It issues $N$ LLM calls, each of which incurs $T_{\text{template}} + T_{\text{context}} + T_{\text{pair}}$ input tokens, where $T_{\text{template}}$ is the prompt template length, $T_{\text{context}}$ the retrieved context length, and $T_{\text{pair}}$ the serialized record-pair length. Thus,
\[
C_{RAG4EM} = N \cdot \bigl(T_{\text{template}} + T_{\text{context}} + T_{\text{pair}}\bigr).
\]
\end{itemize}

In practice, candidate pairs produced by EM blocking often share entities or semantically related records, causing independent retrieval to repeatedly retrieve overlapping context. This redundancy motivates \textit{CE-RAG4EM}, which shares retrieval results and prompt context across similar pairs.

\begin{figure*}[htbp]
  \centering 
  \includegraphics[width=\linewidth]{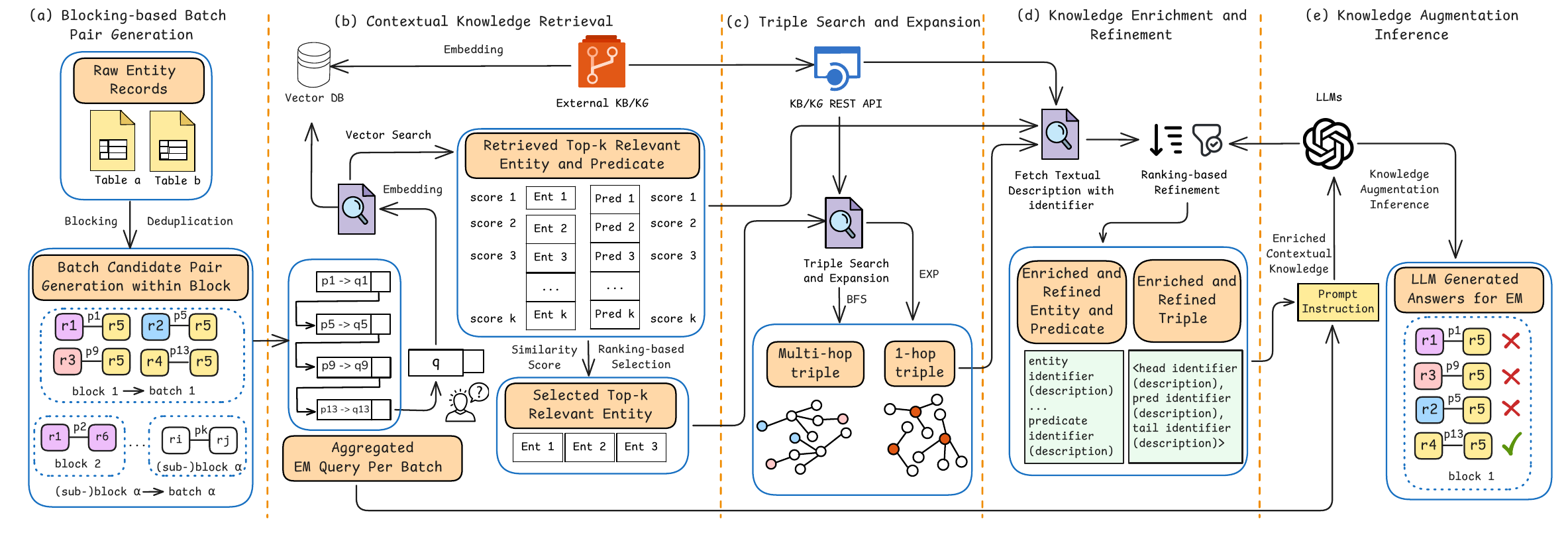}
  \caption{Overview of \textit{CE-RAG4EM}. The framework consists of five phases (a)--(e), detailed in \S\ref{sec:methodology}. A check mark indicates that the LLM predicts \texttt{Yes} for the corresponding EM query, whereas a cross mark indicates \texttt{No}.}
  \label{fig:overall_framework}
\end{figure*}

\section{Methodology}
\label{sec:methodology}

We present \textit{CE-RAG4EM}, a cost-efficient framework for knowledge-grounded entity matching. It addresses two overheads identified in \S~\ref{sec:rag4em} by exploiting an observation specific to the EM pipeline: candidate pairs produced by the same blocking group share blocking characteristics and often exhibit related retrieval needs. \textit{CE-RAG4EM} reuses these blocking-derived groups as batch units for both retrieval and generation. Our design couples batching with the intermediate groups naturally produced by EM blocking. With an average batch size of \(\bar{b}\), the shared overhead can be amortized across approximately \(\bar{b}\) candidate pairs.

Figure~\ref{fig:overall_framework} summarizes \textit{CE-RAG4EM}. The framework first applies blocking over the source and target tables to construct batches of similar candidate pairs (\S\ref{subsec:blocking_pair_generation}). For each batch, the pair queries are aggregated and used to retrieve contextual entities and predicates from the external knowledge graph (\S\ref{subsec:context_retrieval}). The retrieved entities can then be expanded into triples via graph traversal (\S\ref{subsec:triple_search_expansion}). Those entities, predicates, and triples are then enriched with textual descriptions and refined through ranking and instruction-guided filtering (\S\ref{subsec:knowledge_enrich_refine}). Finally, the refined contextual knowledge is injected either into a pair-specific or into a shared batch-level prompt for generation (\S\ref{subse:kag}).

\subsection{Blocking-based Batch Pair Generation}
\label{subsec:blocking_pair_generation}

To reduce retrieval and generation cost, \textit{CE-RAG4EM} organizes candidate pairs into batches using blocking. Let $T_s$ and $T_t$ denote the source and target tables. The construction consists of three steps.

\smallskip

\noindent\textit{Record preprocessing and block generation.}
We first form the unified record space $\mathcal{R} = T_s \cup T_t$ and apply a blocking function~\cite{PKPN14} to group $\mathcal{R}$ into a set of blocks $\mathcal{B} = \{B_1, B_2, \dots, B_\alpha\}$.

\smallskip

\noindent\textit{Candidate pair generation.}
For each block $B$, we generate the candidate pair set
\[
P_B = \{(r_i, r_j) \mid r_i \in (B \cap T_s),\; r_j \in (B \cap T_t)\},
\]

by taking the Cartesian product between source and target records within the block. Since common blocking methods like QGram may assign an entity to multiple blocks based on different blocking keys, two entities can co-occur in more than one overlapping block, causing the same candidate pair to be generated multiple times.

\smallskip

\noindent\textit{Deduplication.}
We retain only the first occurrence of each candidate pair and discard subsequent duplicates. This ensures that each pair is assigned to exactly one block and prevents repeated downstream processing. 
While the design of advanced blocking methods is beyond the scope of this paper, we explore widely used basic blocking methods.
After this step, each block forms a coherent batch of similar candidate pairs and serves as the basic processing unit.

\subsection{Contextual Knowledge Retrieval}
\label{subsec:context_retrieval}

This component follows the same retrieval introduced in \S\ref{subsec:rag4em_pipeline}, but changes the unit of retrieval query from individual pairs to batches.

\noindent\textit{Entity and predicate retrieval.}
Given an entity matching query $q$, a hybrid retrieval that combines dense retrieval and keyword-based search is employed to obtain the Top-$k$ relevant contextual entities $V_k$ and predicates $P_k$ from the external knowledge graph $\mathcal{G}$. For dense retrieval, a pre-trained encoder (Jina Embeddings V3~\cite{SturuaJina2025}) maps both EM queries and KG items into a shared $d$-dimensional embedding space, and a dot product similarity score between the given query and its corresponding entities and predicates is calculated. 
The ranked results from dense and keyword retrieval are then fused using \textit{Reciprocal Rank Fusion (RRF)}, from which the Top-$k$ entities $V_k$ and predicates $P_k$ are selected for knowledge grounding.

\noindent\textit{Blocking-based batch retrieval.}
Instead of issuing one retrieval call per pair, \textit{CE-RAG4EM} retrieves once per batch. If a batch contains $b$ matching pairs, batch retrieval replaces $b$ independent calls with a single aggregated call, and the retrieved knowledge is shared across all pairs in the batch.

\noindent\underline{(a) Threshold-based block decomposition.}
To avoid oversized block groups producing long retrieval queries or batch inference prompts, we impose a maximum batch size \(b_{\max}\). Any blocking-derived candidate group containing more than this number of pairs is split into non-overlapping batches with at most \(b_{\max}\) pairs while preserving the original candidate-pair order in the dataset. The resulting sub-blocks form the final batches for downstream processing.

\noindent\underline{(b) Batch query aggregation.}
Within each batch, we concatenate the queries of all candidate pairs into a single aggregated retrieval query. Because the pairs in the same batch are already semantically similar, the aggregated query remains coherent and typically retrieves context that is broadly useful to the entire batch.

Under this design, batch sizes naturally vary with the sizes of the blocks produced by the blocking method, while an upper bound \(b_{\max}\) is applied to split oversized blocks. This preserves the data-adaptive nature of blocking while keeping retrieval and generation costs manageable.

\subsection{Triple Search and Expansion}
\label{subsec:triple_search_expansion}

After initial retrieval, we further exploit the structure of $\mathcal{G}$ through graph-based expansion, where the Top-$k$ retrieved entities $V_k$ are expanded into triples using two complementary search strategies.

\smallskip

\noindent\textit{Breadth-first search (BFS).}
We use the retrieved entities as starting points and form all source-destination pairs among them. For each pair, BFS is performed over $\mathcal{G}$ to identify multi-hop triples that connect the two entities. The search depth is bounded by $D_{\max}$ to balance search cost and preserve informative structural knowledge.

\smallskip

\noindent\textit{Neighborhood-based expansion (EXP).}
For each entity in $V_k$, EXP retrieves its directly connected neighbors and associated predicates, yielding one-hop structural context~\cite{CaoGLXZX25}. Compared with BFS, EXP emphasizes local context rather than multi-hop connectivity.

\subsection{Knowledge Enrichment and Refinement}
\label{subsec:knowledge_enrich_refine}

The retrieved entities, predicates, and triples must be transformed into textual context before being passed to the LLM.

\noindent\textit{Knowledge enrichment.}
The outputs of vector retrieval and triple expansion are initially symbolic identifiers. We enrich each identifier with its textual description/label from the external KG. Thus, retrieved entities and predicates are represented as \texttt{identifier (description)}, and triples are serialized as \texttt{<head identifier (description), predicate identifier (description), tail identifier (description)>}.

\noindent\textit{Knowledge refinement.}
Although Top-$k$ retrieval already provides a ranked candidate set, it may still contain noisy or weakly relevant context, especially after graph expansion. We therefore apply two refinement mechanisms to make it concise and non-trivial.

\noindent\underline{(a) Vector-similarity ranking-based refinement.}
Entities and predicates are retained in descending order by vector similarity. EXP triples inherit the ranking of their seed entities. BFS triples are selected according to the traversal order induced by the ranked seed entities and the BFS queue.

\noindent\underline{(b) Explicit instruction-guided refinement.}
We further instruct the LLM to assess the relevance of the provided knowledge before inference and to use only helpful context during generation. This reduces the impact of noisy retrieved content without introducing an additional standalone filtering stage, and encourages that only highly relevant contextual information is incorporated into the LLM's reasoning.

\subsection{Knowledge-Augmented Generation}
\label{subse:kag}

After retrieval, triple expansion, enrichment, and ranking-based refinement, the resulting context is incorporated into the prompt. Due to space constraints, we provide the full prompt templates in Appendix~\ref{app:prompt}. \textit{CE-RAG4EM} supports two generation modes.

\noindent\textit{Per-query generation.}
Each candidate pair is combined with the refined context retrieved for its batch and processed through a pair-specific prompt. The prompt instructs the model to first assess the usefulness of the retrieved evidence and then reason over the two records before producing a binary matching decision.

\noindent\textit{Blocking-based batch generation.}
To further reduce token and invocation overhead, all candidate pairs in a batch are serialized into a single prompt with a shared instruction template and shared retrieved context. The LLM then predicts all pairs in one call, amortizing prompt overhead and avoiding repeated context across pairs.

\subsection{End-to-End Solution}
\label{subsec:end-to-end}

Algorithm~\ref{alg:cerag4em} summarizes \textit{CE-RAG4EM}. Given source and target tables $T_s$ and $T_t$, an external KG $\mathcal{G}$, and an LLM $\mathcal{F}_{\text{LLM}}$, the framework proceeds in six stages: blocking, batch candidate pair generation, batch retrieval, optional triple search and expansion, knowledge enrichment and refinement, and batch generation.
\begin{algorithm}[htbp]
\caption{End-to-End \textit{CE-RAG4EM}}
\label{alg:cerag4em}
\begin{algorithmic}[1]
\small
\State \noindent \textbf{Input:} Tables with entity records to be matched $T_s, T_t$, KG $\mathcal{G}$, LLM $\mathcal{F}_{LLM}$, max batch size $b_{\max}$, max triple search depth $D_{\max}$, retrieval budget $k$ 
\State \noindent \textbf{Output:}  Answer for the given EM query $\mathcal{A}$

\State \noindent \textbf{Initialization:} $\mathcal{R} \gets T_s \cup T_t$, $\mathcal{B}, \mathbb{B} \gets \emptyset$, $P_B, \mathcal{Q}_P, \mathcal{G}_P \gets \emptyset$, $\mathcal{A} \gets \emptyset$, $\mathcal{S} \gets \emptyset$

\State $\mathcal{B} \gets \Call{ApplyBlocking}{T_s, T_t}$  
\For{\textbf{each} block $B \in \mathcal{B}$} 
   \State $(P_B, P_{id}) | P_B \in \mathcal{R}  \gets \Call{GenerateMatchingPairs}{B, T_s, T_t}$ 
   
   \State $(P_B, P_{id})\gets \Call{Deduplication}{P_B, P_{id}, \mathcal{S}}$
   \State $\mathcal{S} \leftarrow \mathcal{S} \cup P_B$ 
   \State \textbf{if} $P_B = \emptyset$ \textbf{then continue}
   
   \If{$|P_B| > b_{\max}$}
        \State $\mathbb{B}_{sub} \gets \Call{BatchDecomposer}{P_B, b_{\max}}$ 
        
    \Else
        \State $\mathbb{B}_{sub} \gets \{(P_B, P_{id})\}$
   \EndIf
        \State $\mathbb{B} \gets \mathbb{B} \cup \mathbb{B}_{sub}$
\EndFor
\For{\textbf{each} batch $(P_B, P_{id}) \in \mathbb{B}$}    
    \State $\mathcal{Q}_P \gets \Call{AggregateQueries}{P_B}$
    
    \State $V_k, P_k  \gets \Call{BatchRetrieval}{\mathcal{Q}_P, \mathcal{G}, k}$
    \State $\mathcal{G}_P \gets \Call{TripleSearch}{V_k, \mathcal{G}, D_{\max}}$ 
    \State $\mathcal{G}_P \gets \Call{EnrichRefinement}{\mathcal{G}_P, \mathcal{G}}$ 
    
    \State $ bx \gets \Call{Serializer}{P_B, \mathcal{G}_{P}}$
    \State $\mathcal{A} \gets \mathcal{A} \cup \mathcal{F}_{LLM}.\Call{Inference}{bx}$
    
\EndFor
\State \Return $\mathcal{A}$
\end{algorithmic}
\end{algorithm}
Overall, the design space of \textit{CE-RAG4EM} is defined by two orthogonal dimensions:
\begin{itemize}[leftmargin=*]
    \item \textbf{Blocking-based Optimization (BO):} batch retrieval and batch generation, enabled independently or jointly.
    \item \textbf{Retrieval Granularity (RG):} entity--predicate textual knowledge or triple-based structured knowledge.
\end{itemize}

These choices yield the representative solutions summarized in Table~\ref{tab:design_solution}. While all variants share the same core retrieval, enrichment, and refinement pipeline, \textit{CE-KG-RAG4EM} is distinguished by the additional triple search and expansion phase.

\begin{table}[ht]
\centering
\caption{Representative design solutions in \textit{CE-RAG4EM}.}
\label{tab:design_solution}
\adjustbox{width=1\linewidth}{
\begin{tabular}{cc|cc}
\toprule
\multicolumn{2}{c|}{\multirow{2}{*}{\makecell{Design Solution}}} & \multicolumn{2}{c}{\makecell{Retrieval Granularity (RG)}}               \\ \cline{3-4} 
\multicolumn{2}{c|}{}       & Entity \& Predicate      & Triple             \\ \midrule
\multicolumn{1}{c|}{\multirow{3}{*}{BO}} & \text{\makecell{Batch Retrieval (BR)}} & \textit{CE-RAG4EM-BR}  & \textit{CE-KG-RAG4EM-BR} \\
\multicolumn{1}{c|}{}  & \makecell{Batch Generation (BG)} & \textit{CE-RAG4EM-BG}   & \textit{CE-KG-RAG4EM-BG}    \\
\multicolumn{1}{c|}{}  & \text{\makecell{BR \& BG}}    & \textit{CE-RAG4EM-BR-BG}  & \textit{CE-KG-RAG4EM-BR-BG} \\ \bottomrule
\end{tabular}}

\end{table}

\subsection{Theoretical Analysis}
\label{subsec:theory}
We analyze the \textit{CE-RAG4EM} framework theoretically along two complementary axes: computational cost and matching quality.

\subsubsection{Complexity and Cost Analysis}
\label{subsubsec:complexity}

Let $N$ be the total number of candidate pairs and let $\bar{b}$ be the average batch size after blocking and decomposition. Compared with per-query processing, which requires $N$ retrieval and LLM calls, \textit{CE-RAG4EM} reduces both to approximately $\lceil N/\bar{b}\rceil$ through batch retrieval and generation.

In the analysis below, we use token cost to denote \emph{input} token cost only, since output-token cost depends on the LLM's generation behavior and is reported empirically in \S\ref{subsec:B}. A per-query pipeline incurs
\[
C_{\text{RAG4EM}} = N \cdot (T_{\text{template}} + T_{\text{context}} + T_{\text{pair}}),
\]
whereas \textit{CE-RAG4EM} incurs
\[
C_{\text{CE-RAG4EM}} = \left\lceil N / \bar{b} \right\rceil \cdot (T_{\text{template}} + T_{\text{context}}) + N \cdot T_{\text{pair}}.
\]
Batching amortizes the shared template and context while preserving the pair-specific input cost. 

\begin{proposition}[End-to-end complexity of \textit{CE-RAG4EM}]
\label{prop:complexity}
Let $d_{\max}$ denote the maximum node degree in $\mathcal{G}$, $D_{\max}$ the BFS depth bound, $k$ the per-batch retrieval budget, and $\kappa$ the maximum number of enriched items. For each batch, neighborhood expansion costs $O(kd_{\max})$; and BFS over all pairs of retrieved seed entities has worst-case cost $O(k^2D_{\max}^{d_{\max})}$, while enrichment costs $O(\kappa)$. Over all $N$ candidate pairs, retrieval and LLM generation require $\Theta(\lceil N/\bar{b}\rceil)$ calls.
\end{proposition}

\subsubsection{Matching Quality versus Batch Size}
\label{subsubsec:acc_bs}

For simplicity of exposition, throughout this analysis, we use \(B\) to denote a batch of candidate pairs and \(b=|B|\) to denote its batch size.
Batch retrieval shares a single evidence context $C_B$ across all candidate pairs in a batch. Under a fixed retrieval budget, increasing the batch size may reduce pair-specific relevant evidence while introducing evidence that is relevant to other pairs but irrelevant to the current one. We characterize this effect through evidence coverage and noise.

For a candidate pair $p$, let $R_p^\star$ denote the evidence relevant to its matching decision. Given the shared context $C_B$ of batch $B$, we define
\[
\mathrm{cov}(p,B)
=
\frac{|C_B\cap R_p^\star|}{|R_p^\star|},
\qquad
\mathrm{noise}(p,B)
=
\frac{|C_B\setminus R_p^\star|}{|C_B|}.
\]
Coverage measures how much relevant evidence is retained, whereas noise captures how much of the shared context is irrelevant to the current pair.

We assume that batch-induced coverage degradation and noise increase admit non-decreasing upper bounds
$\delta_{\mathrm{cov}}(b)$ and $\delta_{\mathrm{noise}}(b)$ as functions of batch size $b$, with $\delta_{\mathrm{cov}}(1) = \delta_{\mathrm{noise}}(1) = 0.$ These functions characterize worst-case degradation bounds rather than implying that every larger batch produces worse evidence.

We further assume that the expected prediction error $\mathrm{err}(c,n)$ is non-increasing in coverage $c$, non-decreasing in noise $n$, and Lipschitz continuous:
\[
\left|
\mathrm{err}(c_1,n_1)
-
\mathrm{err}(c_2,n_2)
\right|
\leq
L|c_1-c_2|
+
L'|n_1-n_2|,
\]
for constants $L,L'\geq0$. This captures the behavior that additional relevant evidence generally improves matching reliability, whereas irrelevant evidence may introduce ambiguity. It also assumes that changes in evidence quality induce only bounded changes in classification error.

\begin{theorem}[Batch-size-dependent error bound]
\label{thm:acc_bs}
Under the above assumptions, the average expected prediction-error increase of batch retrieval, relative to per-query retrieval, is bounded by
\[
\frac{1}{b}
\sum_{p\in B}
\left[
\mathrm{err}_{\mathrm{batch}}(p)
-
\mathrm{err}_{\mathrm{pq}}(p)
\right]
\leq
L\delta_{\mathrm{cov}}(b)
+
L'\delta_{\mathrm{noise}}(b)
=:g(b),
\]
where $g(b)$ is non-decreasing in $b$ and satisfies $g(1)=0$.
\end{theorem}

The theorem provides a conditional worst-case characterization of the accuracy--batch-size trade-off: larger batches may improve computational efficiency, while their potential quality degradation is controlled by how rapidly relevant-evidence coverage decreases and retrieval noise increases. 
We further empirically examine this mechanism across different batch sizes. We provide the detailed definitions, intermediate lemma, proof, and the results in Appendix~\ref{app:proof} and Appendix~\ref{app:batch_theory_validation}.

\raggedbottom

\section{Experiments}
\label{sec:experiments}

We present an empirical evaluation of \textit{CE-RAG4EM}.
The codebase is available on \textbf{GitHub.}\footnote{\url{https://github.com/machuangtao/CE-RAG4EM}} 
Our evaluation consists of four parts. 
\textbf{Part A} examines whether external knowledge can mitigate the grounding gap in LLM-based EM, particularly on datasets where LLM-based matchers exhibit high precision but low recall. \textbf{Part B} evaluates whether blocking-aware batch retrieval and batch generation reduce retrieval and inference costs while preserving matching quality. \textbf{Part C} explores the \textit{CE-RAG4EM} design space, focusing on the trade-offs induced by retrieval granularity and blocking strategy. \textbf{Part D} quantifies the impact of batch size on matching quality and computational cost. Additional analyses (KG-RAG design choices, exploration of the number of contexts, scalability across KG domains) are given in Appendix~\ref{app:expb}--\ref{app:expc}.
The main experiments are repeated with three random seeds, and averaged results across three runs are reported. 

\subsection{Experimental Setup}
Unless stated otherwise, all experiments are run on an Ubuntu 22.04 server with 60 CPU cores (Intel Xeon Ice Lake, 2.8GHz), two NVIDIA A100 GPUs (40GB each), and 100GB RAM.

\noindent\underline{Datasets.}
We evaluate \textit{CE-RAG4EM} on nine widely used entity matching benchmarks from the Magellan~\cite{doan2020magellan} and Web Data Commons (WDC)~\cite{primpeli2019wdc}. These datasets span diverse domains and vary in schema complexity, attribute types, and class imbalance. Table~\ref{tab:datasets} summarizes their key statistics. Although some of these datasets are potentially linearly separable in~\cite{10598022} under supervised learning, our evaluation focuses on zero-shot prompting without involving any supervised fine-tuning.

\begin{table}[tb!]
  \centering
  \caption{Nine datasets \cite{doan2020magellan,primpeli2019wdc}, grouped by domain, with summary statistics. Attribute types: T (text), N (numeric), C (categorical), D (date), M (mixed text+numeric); counts in parentheses. \#Pos/\#Neg denote matched/unmatched pairs.}
  \label{tab:datasets}
\adjustbox{width=1\linewidth}{
\begin{tabular}{l l l | r r r r}
\toprule
& \textbf{Dataset} & \textbf{Domain} & \textbf{\#Attr.} & \textbf{\#Attr. Type} & \textbf{\#Pos.} & \textbf{\#Neg.}\\
\midrule
\textsf{ABT} & Abt-Buy & web product & 3 & T(1), N(1), M(1) & 1,028	& 8,547\\
\textsf{AMGO} & Amazon-Google & software & 3 & M(1), T(1), N(1) & 1,167 & 10,293 \\
\textsf{BEER} & Beer & drink & 4 & T(3), N(1) & 68 & 382 \\
\textsf{DBAC} & DBLP-ACM & citation & 4 & T(2), C(1), D(1) & 2,220 & 10,143 \\
\textsf{DBGO} & DBLP-Google & citation & 4 &  T(2), C(1), D(1) & 5,347 & 23,360 \\
\textsf{FOZA} & Fodors-Zagats & restaurant & 6 & T(3), C(1), N(2) & 110 & 836 \\
\textsf{ITAM} & iTunes-Amazon & music & 8 & T(4), N(2), C(1), D(1) & 132 & 407 \\
\textsf{WAAM} & Walmart-Amazon & electronics & 5 & T(1), C(2), N(1), M(1) & 962 & 9,280 \\
\textsf{WDC} & Web Data Commons & web product & 5 & C(2), M(1), T(1), N(1) & 2,250 & 7,992 \\
\bottomrule
\end{tabular}}
\end{table}

\noindent\underline{Knowledge Graph.}
We primarily use Wikidata~\cite{WikidataVrandecicPK23, Difallah2025wikirag} as the external knowledge graph for retrieval and augmentation. Our goal in selecting Wikidata is to provide a \emph{domain-agnostic} and \emph{public} knowledge source that can be applied uniformly across all nine benchmarks, enabling a more universal and reproducible comparison of retrieval strategies. Though domain-specific knowledge graphs (e.g., product or bibliographic KGs) could offer higher coverage and yield stronger results in particular domains, they are not consistently available across datasets and would introduce confounding factors tied to domain engineering. By using Wikidata, we can isolate the impact of \textit{CE-RAG4EM}'s design choices under commonsense knowledge, rather than attributing performance differences to the availability or quality of a domain-specific KG.

Additionally, we examine the scalability of our approach from commonsense knowledge bases to domain-specific knowledge bases by adopting the KG20C~\cite{tran2025kg20c} knowledge graph in the scholarly and citation domain that is derived from the Microsoft Academic Graph. 

\noindent\underline{Knowledge Retrieval.}
For knowledge retrieval from wikidata, we use the public vector index released by the Wikidata Embedding Project.\footnote{\url{https://www.wikidata.org/wiki/Wikidata:Embedding_Project}}
Retrieval is conducted in \emph{natural language}: for each candidate entity pair, we construct a textual query from their attribute descriptions and send it to the embedding service, which returns the nearest Wikidata entities and/or predicates in the embedding space (depending on the retrieval setting).
We further use the Wikidata REST API\footnote{\url{https://www.wikidata.org/wiki/Wikidata:REST_API}} to resolve the retrieved entity/predicate identifiers to labels and descriptions, and to fetch the connected items and relations to construct triples for \textit{CE-KG-RAG4EM}.
For knowledge retrieval from KG20C, we retrieve the relevant entities from KG20C via vector-based retrieval, where the embedding model, vector database, and dense retrieval remain the same as the Wikidata vector database. Then, the triple search and expansion are employed to locally search for triples that are connected to the Top-$k$ relevant entities from dense search and ranking.

\noindent\underline{Evaluation Metrics.}
We evaluate \textit{CE-RAG4EM} along three axes: \textbf{(i) matching quality}, reported primarily with \texttt{F1} and accompanied by \texttt{Precision/Recall}; \textbf{(ii) efficiency}, measured by \texttt{wall-clock latency} (reported per entity pair) and, where informative, separated into retrieval, enrichment, and generation contributions; and \textbf{(iii) cost-related indicators}, where applicable, captured via retrieval workload, e.g., number of retrieval calls (\texttt{RACs}) to reflect blocking-induced overhead and amortization. 
Meanwhile, the input token (\texttt{IT}) and output token (\texttt{OT}), and estimated monetary cost are used to evaluate the cost of the LLM in the \textit{CE-RAG4EM} across different optimization strategies.

\noindent \underline{Blocking Methods.}
We implement blocking using \texttt{pyJedAI}~\cite{Nikoletos0K22}, a widely used open-source EM toolkit, to ensure a reproducible implementation. Full implementation details are provided in Appendix~\ref{app:impl}.
\textsc{QGram} blocking is adopted as the default blocking method because it is a common and robust blocking technique for noisy textual attributes (e.g., typos and lexical variation), which are prevalent in EM benchmarks~\cite{papa2021block}. To assess robustness to the choice of blocking, we further consider two widely used alternatives: \textit{Standard Blocking (StdBlck)} and \textit{Extended QGram blocking (XQGram)}~\cite{papa2021block, MoslemiBM24}. 
The \textit{CE-RAG4EM} can also incorporate the advanced blocking techniques.

\noindent \underline{Backbone LLMs.}
We evaluate both commercial and open-source LLM backbones. We focus on lightweight models that are explicitly positioned as cost-efficient and low-latency in their respective ecosystems. For commercial LLMs, we consider GPT-4o-mini and Gemini~2.0~Flash-Lite, which are designed for cost-efficient inference and fast response times; these models are accessed via the OpenAI API and Google’s Gemini AI APIs. For open-source LLMs, we evaluate Qwen3-4B and Qwen3-8B from the Qwen3 family~\cite{qwen3technicalreport}, deployed locally using vLLM~\cite{kwon2023efficient} for efficient inference. Unless otherwise stated, GPT-4o-mini is used as the default backbone throughout the experiments.

\noindent \underline{Hyperparameters.}
We use standard decoding controls for all LLMs. Temperature and nucleus sampling ($\text{top-}p$) regulate randomness in generation, while $\text{top-}k$ limits sampling to the $k$ most probable next tokens; together they control the determinism and diversity trade-off of model outputs. We cap the maximum generation length at 1024 tokens for all LLMs to bound cost and latency.
The decoding hyperparameters are tuned on \textsf{AMGO} under the LLM-EM setting and reused across all datasets and experiments (temperature $=0.5$, $\text{top-}p=0.8$, $\text{top-}k=20$). 
For Qwen models, we follow the official recommended decoding setup (temperature $=0.7$, $\text{top-}p=0.8$, $\text{top-}k=20$).
We set \(b_{\max}=6\) as the default block size. Accordingly, the batch size remains dynamic and adapts to the block structure of each dataset and has a capped limit of six candidate pairs per batch.
We report the average batch size and block-splitting fraction per dataset in Appendix~\ref{app:block_split_batch_size}. We vary the value of \(b_{\max}\) in \S\ref{subsec:D} to examine its impact on matching quality and computational cost/latency.

\subsection{Part A: The Effectiveness of Grounding Knowledge for Entity Matching}
\label{subsec:A}

\subsubsection{Overall Effectiveness}
\textit{How does \textit{CE-RAG4EM} compare to (i) LLM-only prompting and (ii) supervised PLM-based EM in terms of matching performance, under a zero-shot assumption?}

\noindent \textbf{Evaluation Protocol.}
We compare \textit{CE-RAG4EM} against two baseline families: (i) LLM-EM, which simply prompts the LLM using the input record pair without extra context; and (ii) PLM-based EM, including Ditto~\cite{LiDeepEM20} and Unicorn~\cite{TuFTWL0JG23}.
For \textit{CE-RAG4EM}, we evaluate different retrieval granularities among entity, predicate, and triple. Since our study uses a single predefined test partition, we report \textit{CE-RAG4EM (best-of)} in this experiment: for each dataset, we report the best-performing variant among these configurations as an upper envelope of achievable effectiveness. We use it only for this experiment; the remainder of the paper reports fixed configurations and per-factor analyses.

For Ditto and Unicorn, we use their official implementations and follow the \emph{leave-one-dataset-out} protocol of~\cite{ZhangGCS25}: for each target dataset, the PLM is trained on the remaining datasets and evaluated on the held-out dataset. We adopt this cross-domain setting because the standard supervised in-dataset protocol assumes that labeled data are available for every new domain. In contrast, our target scenario is zero-shot EM with LLMs, where labels may be unavailable or costly to obtain, and training on the evaluation dataset is therefore not possible.

We report \texttt{F1} together with \texttt{Precision} and \texttt{Recall} for effectiveness. For efficiency, we report (a) end-to-end latency per entity pair for \textit{CE-RAG4EM} and LLM-EM, and (b) amortized per-pair PLM training time versus the retrieval and enrichment overhead of \textit{CE-RAG4EM}, where amortized training time is computed as the total training time divided by the number of inference samples.

\noindent\textbf{Results and Analysis.} 
We compare and analyze the results of \textit{CE-RAG4EM} between LLM-EM and PLM-EM, respectively. 

\begin{table}[ht!]
\centering
\caption{F1/Precision/Recall of \textit{CE-RAG4EM} vs. LLM-EM.
Values in parentheses indicate absolute improvements in percentage points.}
\label{tab:cerag4em_llmem_metrics}

\newcommand{\result}[2]{%
  \makebox[2.6em][r]{#1}\,
  \makebox[3.2em][r]{(#2)}%
}
\adjustbox{width=1\linewidth}{
\begin{tabular}{c|cc|cc|cc}
\toprule
\multirow{2}{*}{Dataset} &
\multicolumn{2}{c|}{F1 (\%)} &
\multicolumn{2}{c|}{Precision (\%)} &
\multicolumn{2}{c}{Recall (\%)} \\
\cline{2-7}
& \textit{CE-RAG4EM} & LLM-EM
& \textit{CE-RAG4EM} & LLM-EM
& \textit{CE-RAG4EM} & LLM-EM \\
\midrule

DBGO &
\result{80.77}{+15.75} & 65.02 &
\result{92.45}{-0.68}  & 93.13 &
\result{71.71}{+21.73} & 49.98 \\

ITAM &
\result{72.61}{+8.18} & 64.43 &
\result{97.62}{+0.18} & 97.44 &
\result{58.03}{+9.88} & 48.15 \\

AMGO &
\result{55.47}{+6.82}  & 48.65 &
\result{51.40}{-10.97} & 62.37 &
\result{60.26}{+21.37} & 38.89 \\

FOZA &
\result{83.11}{+7.79} & 75.32 &
\result{100.00}{0.00} & 100.00 &
\result{71.21}{+10.60} & 60.61 \\

BEER &
\result{73.49}{+5.90} & 67.59 &
\result{96.67}{+8.24} & 88.43 &
\result{59.52}{+4.76} & 54.76 \\

DBAC &
\result{81.87}{+5.44} & 76.43 &
\result{95.14}{-1.01} & 96.15 &
\result{71.85}{+8.41} & 63.44 \\

WAAM &
\result{74.85}{+4.13} & 70.72 &
\result{84.59}{-0.97} & 85.56 &
\result{67.18}{+8.29} & 58.89 \\

WDC &
\result{73.35}{+3.58} & 69.77 &
\result{81.74}{-1.73} & 83.47 &
\result{66.53}{+6.60} & 59.93 \\

ABT &
\result{78.72}{+2.47} & 76.25 &
\result{91.27}{-2.20} & 93.47 &
\result{69.26}{+4.66} & 64.60 \\

\bottomrule
\end{tabular}
}
\end{table}

\noindent \underline{\textit{CE-RAG4EM} vs. LLM-EM}. Table~\ref{tab:cerag4em_llmem_metrics} shows that \textit{CE-RAG4EM} outperforms LLM-EM on all nine datasets in terms of \texttt{F1}. 
The gains are primarily recall-driven: across datasets, \textit{CE-RAG4EM} consistently increases recall, indicating that retrieved external evidence helps the LLM identify additional true matches that are missed under direct prompting.

On several datasets (e.g., \textsf{BEER}, \textsf{FOZA}, and \textsf{ITAM}), \textit{CE-RAG4EM} improves \texttt{F1} while maintaining very high precision, suggesting that retrieval provides complementary, relevant signals without introducing many false positives (detailed in \S \ref{app:case_study}). In contrast, on datasets such as \textsf{AMGO}, \textsf{WAAM}, and \textsf{WDC}, we observe a drop in precision relative to LLM-EM even though \texttt{F1} increases. 
A plausible explanation is that retrieved context occasionally includes noisy or weakly related evidence (detailed in failure example 1) in Appendix \ref{appsub:failure-analysis}. 
When record pairs are ambiguous, especially with mixed textual and numeric attributes, such evidence may bias the LLM toward predicting matches, increasing false positives. Overall, these results highlight a precision and recall trade-off inherent to RAG-based EM: retrieving external knowledge can substantially improve coverage (recall), but its benefits depend on the relevance of the retrieved context, motivating the later analysis of retrieval granularity and context construction.

\begin{figure}[t]
  \centering
  \begin{subfigure}[t]{0.9\linewidth}
    \centering
    \includegraphics[width=\linewidth]{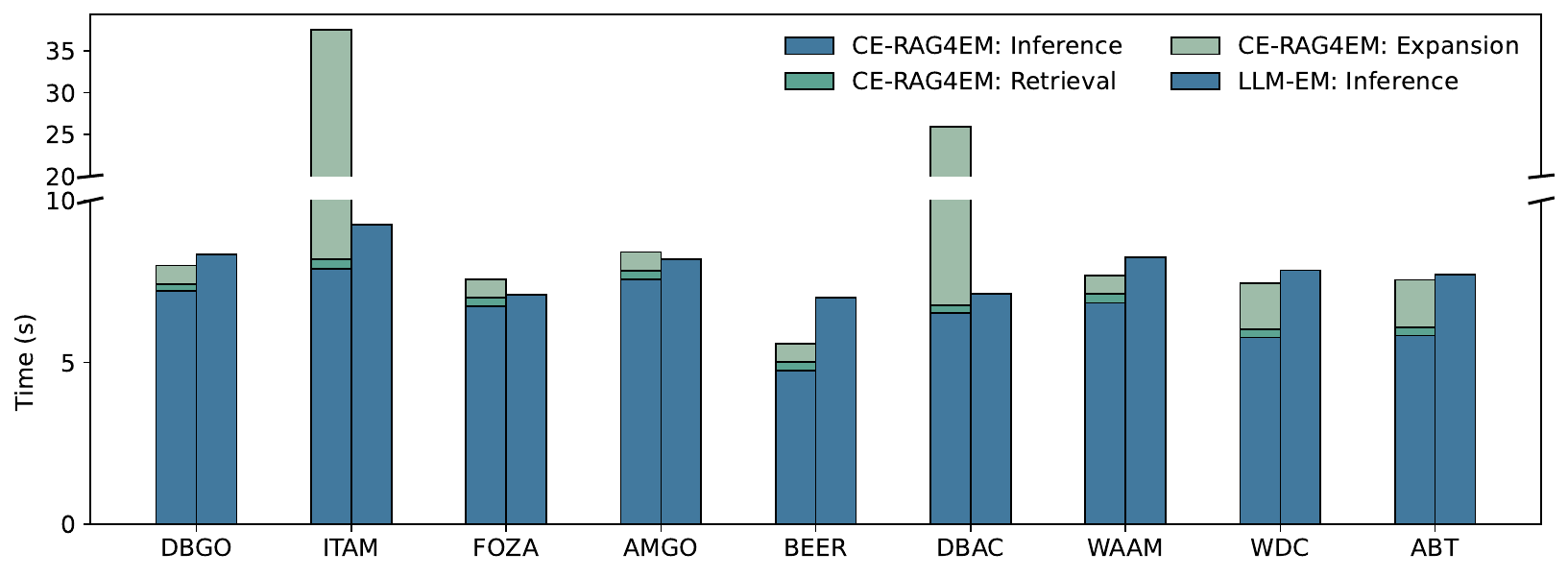}
    \caption{End-to-end matching time per entity pair.}
    \label{fig:efficiency:a}
  \end{subfigure}

  \begin{subfigure}[t]{0.9\linewidth}
    \centering
    \includegraphics[width=\linewidth]{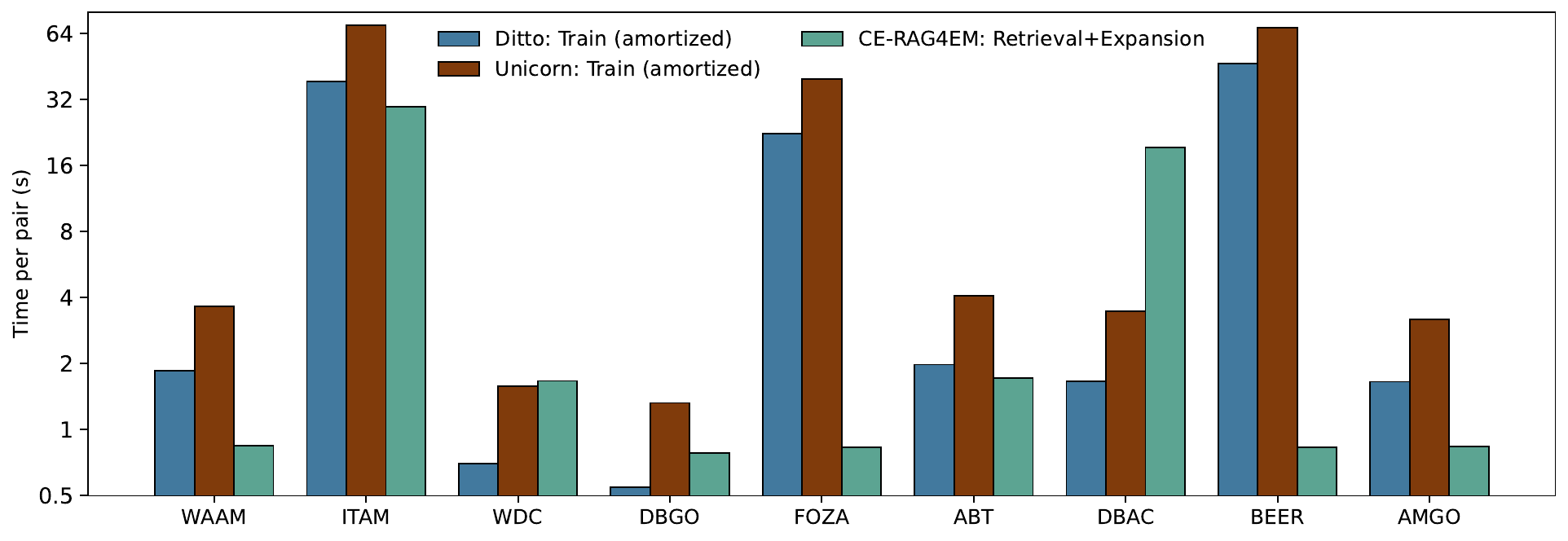}
    \caption{Amortized training time (PLMs) vs. retrieval + expansion time (\textit{CE-RAG4EM}).}
    \label{fig:efficiency:b}
  \end{subfigure}

  \caption{Efficiency comparison of \textit{CE-RAG4EM} against (a) LLM-EM and (b) PLM baselines.}
  \label{fig:efficiency}
\end{figure}

Figure~\ref{fig:efficiency:a} shows that \textit{CE-RAG4EM} often reduces the inference cost relative to LLM-EM. Although retrieval and enrichment add extra overhead, the augmented context typically enables the LLM to generate fewer output tokens, which reduces generation time and cost, and further leads to a net end-to-end speedup. In several datasets (e.g., \textsf{DBGO} and \textsf{BEER}), this inference-time reduction is large enough to offset the additional overhead of retrieval and enrichment, yielding lower overall latency per entity pair.

Retrieval consistently accounts for only a small fraction of the time across datasets, suggesting that vector search is not the primary bottleneck in our setup. In contrast, enrichment time can become substantial on some datasets and may dominate the pipeline, indicating that the current implementation of enrichment (e.g., resolving identifiers and expanding neighborhood information) is a key target for further optimization in later design-space analysis.

\noindent \underline{CE-RAG4EM vs. PLM-EM}.
Table~\ref{tab:cerag4em_ditto_unicorn_metrics} compares \textit{CE-RAG4EM} with supervised PLM baselines (Ditto and Unicorn) under the leave-one-dataset-out protocol.

\begin{table}[ht!]
\centering
\caption{\textit{CE-RAG4EM} vs. PLMs (Ditto, Unicorn). Datasets are sorted by \textit{CE-RAG4EM} advantage.}
\label{tab:cerag4em_ditto_unicorn_metrics}
\renewcommand{\arraystretch}{1.1}
\adjustbox{width=1\linewidth}{
\begin{tabular}{c|ccc|ccc|ccc}
\toprule
\multirow{2}{*}{Dataset} &
\multicolumn{3}{c|}{F1 (\%)} &
\multicolumn{3}{c|}{Precision (\%)} &
\multicolumn{3}{c}{Recall (\%)} \\
\cline{2-10}
& \textit{CE-RAG4EM} & Ditto & Unicorn
& \textit{CE-RAG4EM} & Ditto & Unicorn
& \textit{CE-RAG4EM} & Ditto & Unicorn \\
\midrule
WAAM & 74.85 & 56.50 & 61.47 & 84.59 & 64.32 & 70.99 & 67.18 & 50.37 & 54.20 \\
ITAM & 72.61 & 64.33 & 67.45 & 97.62 & 67.32 & 68.12 & 58.03 & 61.59 & 66.79 \\
WDC  & 73.35 & 45.16 & 70.03 & 81.74 & 49.89 & 70.24 & 66.53 & 41.26 & 69.83 \\
DBGO & 80.77 & 77.62 & 78.06 & 92.45 & 81.23 & 81.71 & 71.71 & 74.31 & 74.72 \\
FOZA & 83.11 & 69.64 & 82.61 & 100.00 & 89.16 & 90.13 & 71.21 & 57.13 & 76.25 \\
ABT  & 78.72 & 61.95 & 78.72 & 91.27 & 64.71 & 89.41 & 69.26 & 59.43 & 70.32 \\
DBAC & 81.87 & 82.96 & 88.72 & 95.14 & 87.65 & 92.12 & 71.85 & 78.76 & 85.57 \\
BEER & 73.49 & 84.19 & 82.20 & 96.67 & 87.33 & 84.39 & 59.52 & 81.27 & 80.13 \\
AMGO & 55.47 & 53.86 & 68.86 & 51.40 & 56.70 & 67.97 & 60.26 & 51.29 & 69.77 \\
\bottomrule
\end{tabular}
}
\end{table}

Overall, \textit{CE-RAG4EM} remains competitive without requiring any data labels: it outperforms PLMs on several datasets (e.g., \textsf{WAAM}, \textsf{ITAM}, \textsf{WDC}) and is comparable on others (e.g., \textsf{DBGO}, \textsf{FOZA}, \textsf{ABT}). This suggests that external knowledge retrieval can partially substitute for target-domain supervision in specific cases.

PLMs are stronger on \textsf{DBAC}, \textsf{BEER}, and \textsf{AMGO}, which clarifies when supervised transfer is advantageous. On \textsf{DBAC}, PLMs likely benefit from transferable schema-level cues that generalize well across citation-style datasets. On \textsf{BEER}, \textit{CE-RAG4EM} achieves very high precision but lower recall, suggesting that retrieval provides strong evidence for only a subset of true matches, leading to conservative match decisions and missed positives when evidence is absent or not retrieved. On \textsf{AMGO}, \textit{CE-RAG4EM} shows lower precision than \textit{Unicorn}, consistent with the hypothesis that ambiguous records and mixed attribute types increase the chance of weakly relevant retrieved context, which can bias the LLM toward over-matching and introduce additional false positives.

Figure~\ref{fig:efficiency:b} compares amortized per-pair PLM training cost (Ditto / Unicorn) with the per-pair retrieval and enrichment overhead of \textit{CE-RAG4EM}. Since PLM inference (locally) is not directly comparable to closed-source API inference, we focus on the per-dataset ``setup'' overhead: PLMs pay an upfront training cost (amortized over test pairs), whereas \textit{CE-RAG4EM} pays a per-pair augmentation cost. A clear trend is that on smaller datasets, the amortized PLM training overhead is large, while \textit{CE-RAG4EM}'s overhead is comparatively stable and is dominated by knowledge enrichment and expansion. Moreover, on six out of nine datasets, such as \textsf{WAAM} and \textsf{ITAM}, the per-pair retrieval+enrichment overhead of \textit{CE-RAG4EM} is lower than the amortized training overhead of Ditto/Unicorn, implying that retrieval-based augmentation can incur less per-dataset overhead than cross-dataset PLM training for label-scarce workloads.

\noindent\textbf{Remark on the PLM comparison.} Ditto and Unicorn are fine-tuned PLMs trained with full access to task-specific labeled entity pairs; they serve as a \emph{cross-paradigm reference point} rather than a direct competitor. Our target setting is fundamentally different: \textit{CE-RAG4EM} requires no labeled data and is applicable in scenarios where annotation is unavailable or costly. The comparison is included to situate our work in the broader EM landscape and to quantify the gap between supervised and unsupervised approaches. That \textit{CE-RAG4EM} outperforms both Ditto and Unicorn on \textsf{WAAM}, \textsf{ITAM}, and \textsf{WDC} without any supervision demonstrates that retrieval-based grounding can partially substitute for labeled supervision in specific settings.

\subsubsection{Comparison with Additional Baselines} \label{sec:addtional_baselines}
\textit{How does \textit{CE-RAG4EM} compare to (i) an \textit{Offline-Enrich} baseline with entity linking tools and (ii) a RAG pipeline with a PLM-based reranking?}

\noindent \textbf{Evaluation Protocol.}
To distinguish hybrid retrieval from simply augmenting records with external KG information, we additionally consider an Offline-Enrich baseline with entity linking tools via falcon2.0\footnote{https://github.com/SDM-TIB/falcon2.0} and the same item enrichment in \textit{CE-RAG4EM}. 
It retrieves Wikidata evidence for each source and target record independently and appends the retrieved information to the corresponding records before LLM-based matching, without performing query-dependent retrieval.
We also compare \textit{CE-RAG4EM} against the RAG+Re-Rank baseline to evaluate whether an explicit reranking stage improves the retrieval pipeline. This baseline follows the same online retrieval procedure as \textit{CE-RAG4EM}, but applies a cross-encoder\footnote{https://huggingface.co/cross-encoder/ms-marco-MiniLM-L4-v2} to rerank the retrieved KG evidence before the knowledge grounding.

\begin{table}[t]
\centering
\caption{F1 comparison of \textit{CE-RAG4EM} with additional offline KG enrichment and explicit reranking baselines.}
\label{tab:retrieve_baselines}
\adjustbox{width=1\linewidth}{
\begin{tabular}{lcccccccccc}
\toprule
Method
& \textsf{ABT} & \textsf{AMGO} & \textsf{BEER} & \textsf{DBAC} & \textsf{DBGO} & \textsf{FOZA} & \textsf{ITAM} & \textsf{WAAM} & \textsf{WDC} & \textbf{AVG.} \\
\midrule
Offline-Enrich
& \textbf{78.90}& \underline{54.08}& \underline{70.62}& \underline{82.95}& 80.34 & 78.88 & 71.65 & 72.50 & 73.26 & 73.69 \\

RAG+Re-Rank
& 78.26 & 53.47 & 65.61 & \textbf{83.62} & \textbf{81.71} & \textbf{83.17} & \underline{71.67}& \underline{73.29}& \textbf{74.37} & \underline{73.91}\\ 

\midrule

CE-RAG4EM
& \underline{78.72}& \textbf{55.47} & \textbf{73.49} & 81.87 & \underline{80.77}& \underline{83.11}& \textbf{72.61} & \textbf{74.85} & \underline{73.35}& \textbf{74.92} \\
\bottomrule
\end{tabular}
}
\end{table}

\noindent \textbf{Results and Analysis.}
We provide the analysis on the effectiveness of our \textit{CE-RAG4EM} against two additional baselines below.

\noindent \underline{\textit{CE-RAG4EM} vs. Offline-Enrich baseline}. As shown in Table~\ref{tab:retrieve_baselines}, 
compared with Offline-Enrich, \textit{CE-RAG4EM} achieves higher F1 on seven of the nine datasets and improves the average F1 from 73.69 to 74.92. 
The relatively strong performance of offline enrichment confirms that access to Wikidata evidence itself contributes substantially to entity matching. 
However, the additional gains from \textit{CE-RAG4EM} indicate that its improvements cannot be attributed solely to static KG augmentation: conditioning retrieval on the matching query provides complementary benefits beyond independently enriching the two input records.

\noindent \underline{\textit{CE-RAG4EM} vs. RAG + Reranking}. 
Table~\ref{tab:retrieve_baselines} also shows that adding an explicit reranker does not produce a consistent improvement over \textit{CE-RAG4EM}. Although RAG+Re-Rank performs better on \textsf{DBAC}, \textsf{DBGO}, and \textsf{WDC} datasets, the margins are small, and its average F1 (73.91) remains below that of \textit{CE-RAG4EM} (74.92). These results suggest that explicit reranking can improve evidence prioritization for some datasets, but is not uniformly beneficial under our setting.
One possible explanation is that \textit{CE-RAG4EM} already incorporates relevance-aware ranking and refinement, leaving 
relatively little room for a reranking to improve the retrieved context.

\subsubsection{Backbone LLM Generalization}

\textit{Do the benefits of \textit{CE-RAG4EM} persist across backbones with different sizes and architectures?}

\noindent \textbf{Evaluation Protocol.}
We evaluate backbone generalization by instantiating the same \textit{CE-RAG4EM-BR} pipeline with the Top-1 \texttt{QID} on four backbone LLMs: GPT-4o-mini, Gemini-2.0-flash-lite, Qwen3-4B, and Qwen3-8B. For each backbone, we measure effectiveness for the \textit{Base} (LLM-EM) setting and the corresponding \textit{CE-RAG4EM} setting, and summarize the effect of retrieval using paired differences $\Delta$F1 $=$ F1(\textit{CE-RAG4EM}) $-$ F1(\textit{Base}).

\begin{figure}[t]
    \centering
    \includegraphics[width=1\linewidth]{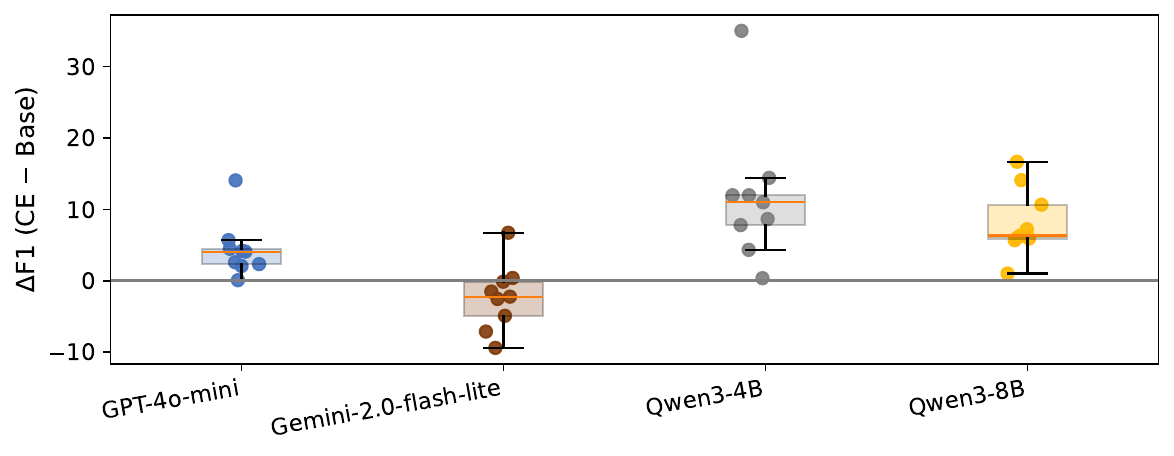}
    \caption{Backbone LLM Generalization.  $\Delta$F1$=$F1(\textit{CE-RAG4EM})$-$F1(\textit{Base}) distribution by LLMs. Dots are datasets; boxplots show median and interquartile range. $\Delta$F1$>0$ indicates improvement by knowledge grounding.}
    \label{fig:exp8_delta_f1}
\end{figure}

\noindent \textbf{Results and Analysis.}
Figure~\ref{fig:exp8_delta_f1} shows that \textit{CE-RAG4EM} yields predominantly positive $\Delta$\texttt{F1} across datasets for GPT-4o-mini and both Qwen3 backbones, indicating that lightweight grounding transfers across architectures. 
The gains are more pronounced and less stable for smaller backbones (notably Qwen3-4B). This pattern is consistent with a \emph{capacity/knowledge} hypothesis~\cite{pan2026understanding, roberts2020much}: smaller models have less parametric knowledge, so they benefit more from explicit, structured evidence that narrows the hypothesis space. As model capacity increases (e.g., Qwen3-8B and GPT-4o-mini), the base setting is already stronger, so retrieval yields smaller but still generally positive improvements, suggesting diminishing returns when the backbone can already infer many matches from surface cues.
Gemini-2.0-flash-lite exhibits higher variance, with $\Delta$\texttt{F1} values closer to zero and occasional degradations. 

One plausible explanation is \emph{evidence utilization}: different model families often differ in how they prioritize external context relative to the input record pair under the same prompt and budget. When the retrieved entity is highly relevant, it helps the model; when it is weakly related (e.g., due to ambiguous mentions or noisy attributes), certain models may over-weight the added context and drift toward incorrect matches,
producing negative $\Delta$F1 on a subset of datasets. This observation again stresses the previous finding that retrieval quality and context relevance are key factors of \textit{RAG4EM}, and suggests backbone-specific prompting or evidence filtering.

Overall, this experiment demonstrates that \textit{CE-RAG4EM} generalizes across backbone models. Smaller open-source models benefit more from external grounding, whereas some lightweight commercial models show high sensitivity to context quality.

\subsection{Part B: Efficiency and Cost Analysis}
\label{subsec:B}
\subsubsection{Time Efficiency}
\textit{How do batch optimizations in \textit{CE-RAG4EM} compare to per-query RAG4EM in terms of end-to-end time?}

\noindent \textbf{Evaluation Protocol.}
We compare three representative RAG-based EM pipelines that differ in how retrieval and generation are executed:
(i) RAG4EM, which performs retrieval and generation independently for each entity pair (per-query);
(ii) \textit{CE-RAG4EM-BR}, which performs \emph{batch retrieval} once per batch and reuses the retrieved context for all pairs in the batch, while generation is still executed per pair; and
(iii) \textit{CE-RAG4EM-BG}, which performs \emph{batch generation} by sending all pairs in a batch as one LLM request, while retrieval is still executed per pair.
For all variants, we keep the retrieval configuration fixed and use the same context budget (Top-$k=2$ retrieved Wikidata entity \texttt{QID}).
We report effectiveness using \texttt{F1}. For efficiency, we report end-to-end \emph{time per pair}. For the batched variants, we attribute shared costs, e.g., a single batch-level retrieval in \textit{CE-RAG4EM-BR} or a single batch-level generation in \textit{CE-RAG4EM-BG} back to individual pairs by dividing the batch-level time by the number of pairs within the batch. We then decompose time into retrieval, expansion, and generation to identify the most beneficial stage from batching.

\begin{figure}[t]
  \centering
  \includegraphics[width=1\linewidth]{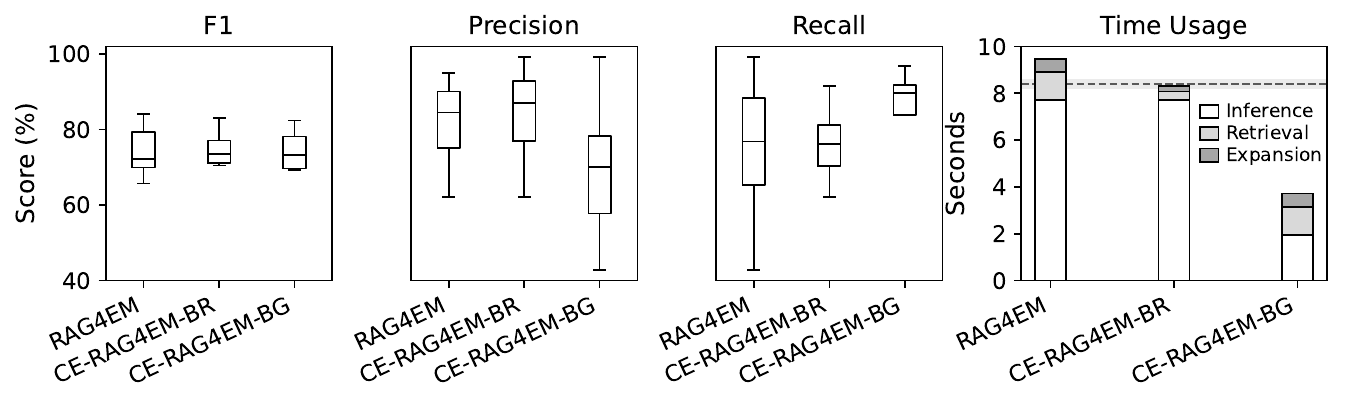}
  \caption{Batching vs.\ per-query. F1/Prec./Rec. and end-to-end time per pair for RAG4EM, \textit{CE-RAG4EM-BR} (batch retrieval), and \textit{CE-RAG4EM-BG} (batch generation). Dashed line in the time plot: LLM-EM runtime.}
  \label{fig:batch_retrieval_delta_f1}
\end{figure}

\noindent \textbf{Results and Analysis}.
Figure~\ref{fig:batch_retrieval_delta_f1} compares per-query RAG4EM with two batching optimizations: \textit{CE-RAG4EM-BR} (batch retrieval) and \textit{CE-RAG4EM-BG} (batch generation). 

Overall, the three variants achieve similar \texttt{F1} on average, indicating that batching primarily changes the \emph{precision--recall balance} and the \emph{system overhead} rather than shifting accuracy uniformly.
\textit{CE-RAG4EM-BG} tends to increase recall but can reduce precision, whereas \textit{CE-RAG4EM-BR} preserves a more stable balance. A plausible explanation is that batch generation presents multiple pairs together in a single prompt, which can introduce \emph{cross-pair coupling}: evidence or patterns from some pairs may influence the model’s decisions on others. This can make the model more prone to predict matches, improving recall (fewer missed positives) but also increasing false positives when weakly related context is inadvertently shared or when the model adopts a more ``match-biased'' decision rule for consistency within the batch. In contrast, \textit{CE-RAG4EM-BR} reuses retrieval within a block but still generates decisions per pair, which avoids LLM coupling across pairs and helps maintain precision while still benefiting from retrieval reuse.
From an efficiency perspective, \textit{CE-RAG4EM-BG} substantially reduces inference time per pair, making it highly competitive despite its weaker precision stability. This reduction is expected because batch generation amortizes fixed LLM invocation overhead across multiple pairs and can shorten total decoding by producing a compact batched output format. These results suggest that batch generation is a promising system optimization, but it requires careful prompt and output design (e.g., stronger per-pair isolation or calibration) to avoid precision degradation while retaining its large runtime advantages.

\subsubsection{Token Cost}
\textit{How do batch optimizations in \textit{CE-RAG4EM} compare to per-query RAG4EM in terms of token cost?}

\noindent \textbf{Evaluation Protocol.} We estimate the LLM cost from token usage, using the input-token (\texttt{IT}) and output-token (\texttt{OT}) charges under GPT-4o-mini pricing. For batched requests, we amortize the average token and monetary cost to the pair level by dividing by the number of pairs in the batch. 

\begin{figure}[t]
    \centering
    \includegraphics[width=1\linewidth]{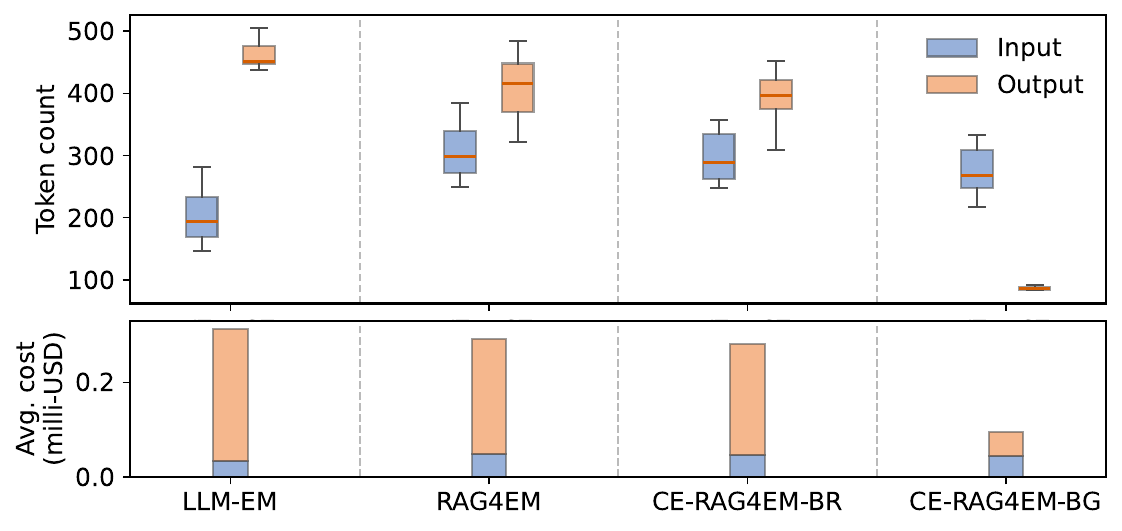}
    \caption{Token cost analysis. Top: distribution of input (\texttt{IT}) and output (\texttt{OT}) tokens per entity pair across four pipeline variants. Bottom: corresponding average cost per pair.}
    \label{fig:token_cost}
\end{figure}

\noindent \textbf{Results and Analysis}.
Figure~\ref{fig:token_cost} extends the efficiency analysis with a token-level cost breakdown.
The top panel shows the per-pair distribution of input tokens (\texttt{IT}) and output tokens (\texttt{OT}) across all four variants; the bottom panel reports the corresponding average monetary cost in milli-USD.

The dominant cost driver across all variants is the \emph{output token count}.
LLM-EM produces the highest \texttt{OT} ($\approx$470 tokens per pair) at the lowest \texttt{IT} ($\approx$200 tokens), as the model must generate verbose, self-reliant justifications without any external grounding.
Introducing retrieval (RAG4EM) raises \texttt{IT} by $\approx$100 tokens due to the added context, and yields a modest reduction in \texttt{OT} (to $\approx$400 tokens): with an external evidence anchor, the model can produce slightly more concise responses.
\textit{CE-RAG4EM-BR} exhibits similar token consumption to RAG4EM, with a slight reduction in input tokens. This is because each pairwise request still relies on the same amount of retrieved information as in RAG4EM, except that the retrieved content is shared among requests within the same batch (the primary cost reduction of \textit{CE-RAG4EM-BR} comes from reducing the number of retrieval calls).

The decisive cost reduction comes from \emph{batch generation}.
\textit{CE-RAG4EM-BG} collapses \texttt{OT} from $\approx$400 to $\approx$90 tokens per pair, cutting total cost to $\approx$0.06 milli-USD, roughly a $4.5\times$ reduction.
This is the token-level manifestation of the above amortization effect: batching multiple pairs into a single LLM request allows the model to produce one compact, structured response (a list of decisions), eliminating the repeated formatting preambles and chain-of-thought scaffolding that inflate per-pair \texttt{OT} in the individual-generation setting.
Together, these results show that the path to cost efficiency in RAG4EM runs through the \emph{output side}: batch generation delivers the largest cost reduction, motivating \textit{CE-RAG4EM-BG} as the preferred configuration in cost-sensitive deployments, subject to the previously identified trade-off.

\subsection{Part C: Design Space Exploration}
\label{subsec:C}

\subsubsection{Retrieval Granularity}
\label{subsec:C:retrieval_granularity}
\textit{How does retrieval granularity affect the effectiveness and overhead of \textit{CE-RAG4EM}: node-level retrieval in \textit{CE-RAG4EM-BR} versus KG triple in \textit{CE-KG-RAG4EM-BR}?}

\noindent \textbf{Evaluation Protocol.}
We study retrieval granularity by comparing:
(i) node-level retrieval in \textit{CE-RAG4EM-BR}, where the retrieved context consists of either Wikidata predicates (PID) or entities (QID); and
(ii) triple-level retrieval in \textit{CE-KG-RAG4EM-BR}, where the context is constructed as a small set of Wikidata triples generated via either expansion-based traversal (EXP) or breadth-first search (BFS) starting from the retrieved nodes.
Across all variants, we control the retrieval budget by using the Top-$k=2$ retrieved nodes to construct the context. We compare effectiveness using \texttt{F1} across datasets. For efficiency, since all variants perform the same Top-$k$ retrieval and have similar token budgets in our setup, we focus on the additional triple construction/enrichment overhead to isolate the cost introduced by KG-based context.

\captionsetup[figure]{skip=2pt} 
\captionsetup[subfigure]{skip=-2pt} 
\captionsetup[subfigure]{font=small}
\begin{figure}[t]
  \centering

  \begin{subfigure}[t]{0.66\linewidth}
    \centering
    \includegraphics[width=1\linewidth]{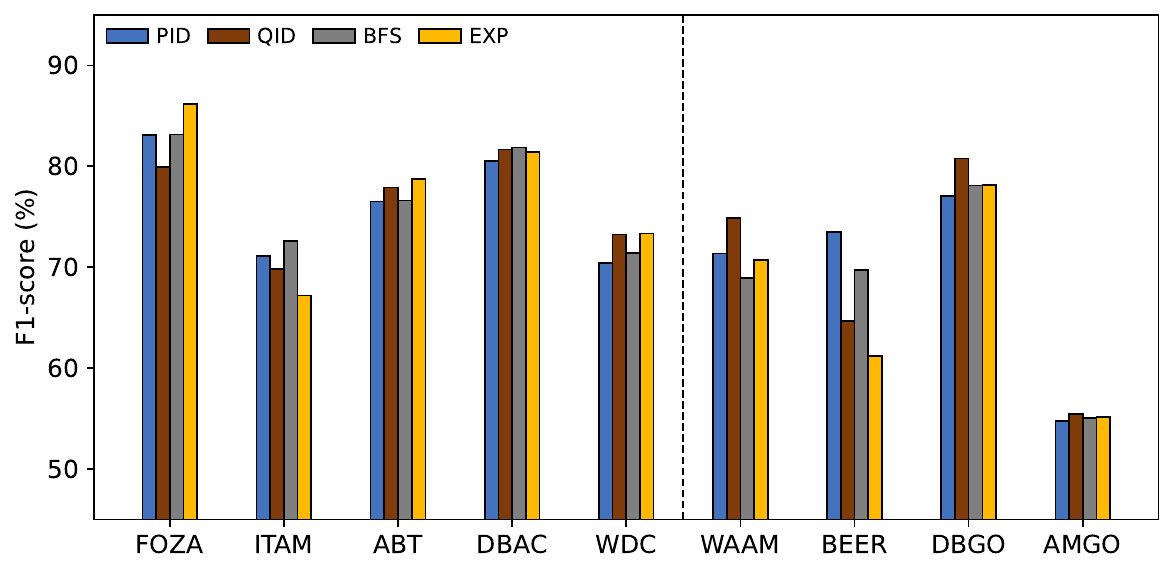}
    \caption{F1 comparison}
    \label{fig:rg-effectiveness-a}
  \end{subfigure}\hfill
  \begin{subfigure}[t]{0.33\linewidth}
    \centering
    \includegraphics[width=1\linewidth]{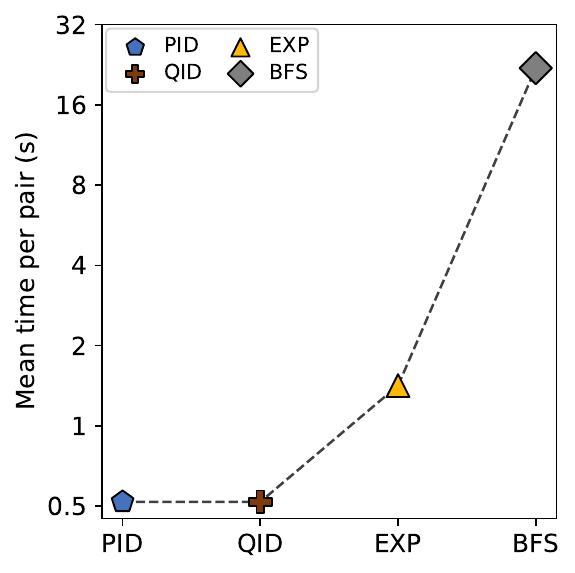}
    \caption{Time comparison}
    \label{fig:rg-efficiency-b}
  \end{subfigure}

  \caption{Retrieval Granularity. PID/QID: node-level retrieval in \textit{CE-RAG4EM-BR}. EXP/BFS: KG-triple context construction in \textit{CE-KG-RAG4EM-BR} via neighboring expansion EXP or BFS. (a) F1 by dataset (sorted by KG-variant advantage). (b) Mean context-construction time per entity pair.}
  \label{fig:rg-efficiency}
\end{figure}

\noindent\textbf{Results and Analysis.}
Figure~\ref{fig:rg-effectiveness-a} compares node-level retrieval (\textit{CE-RAG4EM-BR} with PID/QID) and triple-level retrieval (\textit{CE-KG-RAG4EM-BR} with EXP/BFS). Overall, KG-based triple construction is most beneficial on \textsf{FOZA}, \textsf{ITAM}, \textsf{ABT}, \textsf{DBAC}, and \textsf{WDC}, while node-level retrieval is competitive or better on the remaining datasets. 
Specifically, \textit{CE-RAG4EM} with EXP achieved the best results on \texttt{FOZA} and \texttt{ABT} and comparable results on \texttt{DBAC}, \texttt{WDC}, and \texttt{AMGO}.
These ``KG-friendly'' datasets contain sparse or ambiguous attribute descriptions (e.g., short names, missing identifiers, mixed fields), where a small set of relational triples provides disambiguating context (e.g., type, brand/artist/venue, location), whereas isolated entities/predicates may be insufficient to resolve ambiguity.
This suggests that the overall retrieval granularity selection should shift from node‑level (PID or QID) to triple‑level (EXP) as attribute diversity and ambiguity increase.
Across datasets, EXP is frequently among the top-performing strategies and is the best choice on several datasets. This can be attributed to EXP constructing a smaller and more focused triple around the retrieved nodes compared to BFS, which improves relevance and reduces noise when the initial retrieval is accurate. However, as it relies heavily on the quality of the starting nodes and a limited expansion budget, the gains vary across datasets.

Figure~\ref{fig:rg-efficiency-b} shows the trends of retrieval time. PID and QID incur short retrieval time, while KG-based methods introduce additional triple-search time, with BFS being the most expensive. This difference matches the search behavior: EXP limits expansion to a small, focused neighborhood, whereas BFS explores more broadly and thus incurs higher traversal time. In return, BFS sometimes delivers the largest \texttt{F1} gains, suggesting a quality and cost trade-off: broader traversal improves coverage but risks higher latency and noise, while EXP provides a more stable middle ground.

\subsubsection{Blocking Strategy Robustness}
\label{subsec:C:blocking_strategy}
\textit{How sensitive is \textit{CE-RAG4EM} to the choice of blocking method?}

\begin{figure}[t]
  \centering
  \includegraphics[width=1\linewidth]{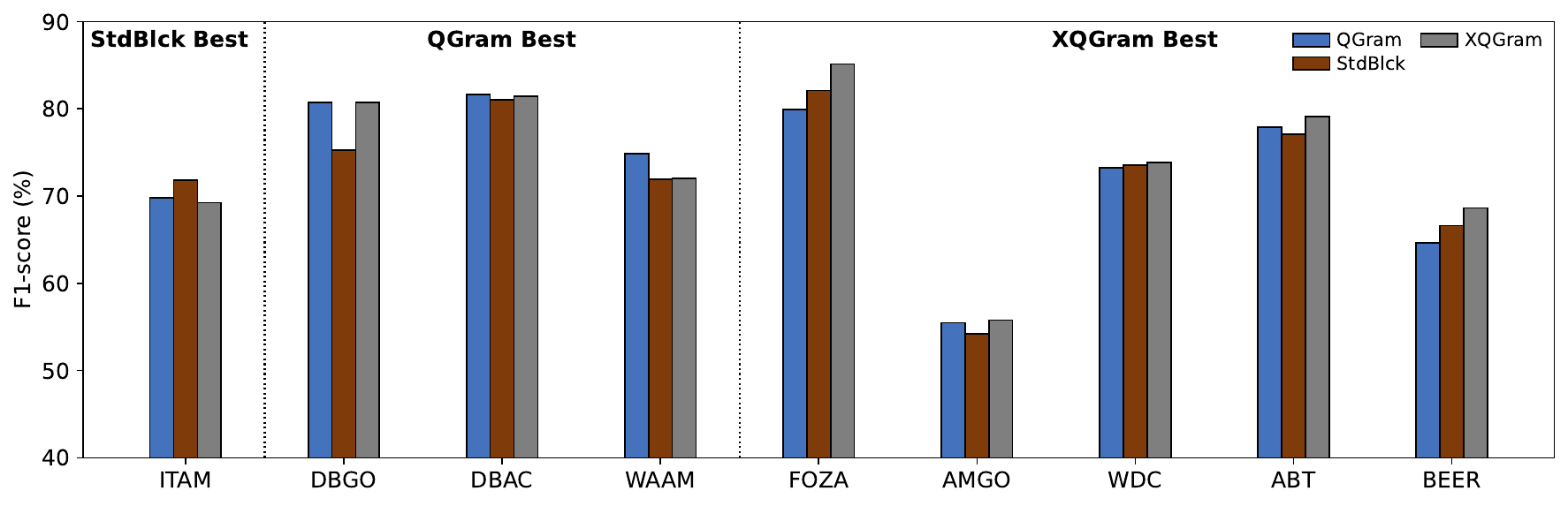}
  \caption{Blocking Robustness. F1 comparison of \textit{CE-RAG4EM-BR} under three blocking methods. Datasets are grouped according to the blocking method that achieves the best performance (separated by dotted lines).}
  \label{fig:blocking_robustness}
\end{figure}

\noindent \textbf{Evaluation Protocol.}
To isolate the impact of blocking, we fix the remainder of the pipeline: all configurations use GPT-4o-mini and \textit{CE-RAG4EM-BR} with Top-$k=2$ retrieved Wikidata items (QID) per block, using the same aggregated query construction, refinement, and prompting template.
For each dataset, we generate blocks using three widely used unsupervised blocking methods: \texttt{StdBlck} (standard token-based), \texttt{QGram} (character $q$-gram), and \texttt{XQGram} (extended $q$-gram).

\noindent \textbf{Results and Analysis.}
Figure~\ref{fig:blocking_robustness} reports the \texttt{F1} of \textit{CE-RAG4EM-BR} under the three blocking methods.  Overall, the resulting matching performance is stable across blocking choices, and $q$-gram based methods (\texttt{QGram}/ \texttt{XQGram}) outperform \texttt{StdBlck} on most datasets. The reason is that many benchmark datasets contain noisy textual fields and formatting variation (Table~\ref{tab:datasets}); exact token-based blocking can fragment near-duplicates into different blocks, while $q$-gram signatures tolerate typos and lexical variation and thus produce more coherent batches for further retrieval.

Among the $q$-gram methods, \texttt{XQGram} often achieves the best or near-best matching quality, whereas \texttt{QGram} remains competitive with consistently lower blocking cost. This makes it a better overall operating point, so we use it as the default in the main experiments. Since \textit{CE-RAG4EM-BR} performs retrieval at the batch level, the blocking method must also balance block purity and coverage: overly broad blocks may mix heterogeneous pairs and introduce less relevant evidence, whereas overly fragmented blocks reduce opportunities for related pairs to share retrieved context. In our experiments, \texttt{QGram} shows a favorable balance between these factors, offering competitive matching quality with lower blocking overhead across diverse datasets.

\subsection{Part D: Batch Size Sensitivity} 
\label{subsec:D}

\noindent\textbf{Research Question.}
\textit{How does batch size impact the effectiveness--efficiency trade-off in \textit{CE-RAG4EM}?}

\noindent \textbf{Evaluation Protocol.}
We evaluate the sensitivity of the maximum batch size $b_{\max}$ that controls the upper bound on the number of candidate pairs processed per batch and thus affects batch retrieval in \textit{CE-RAG4EM-BR}.
For each dataset, we run \textit{CE-RAG4EM-BR} with \texttt{QGram} blocking while varying $b_{\max} \in \{2,4,6,8\}$. The retrieved context consists of the Top-$2$ Wikidata items (QID). When a raw block exceeds $b_{\max}$, we apply the same threshold-based decomposition strategy as in \S\ref{subsec:context_retrieval} to split it into non-overlapping sub-blocks.
We quantify the (i) matching quality measured by \texttt{F1}, and (ii) retrieval cost measured by the number of retrieval API calls (\texttt{RACs}).

\begin{figure}[ht]
  \centering
  \includegraphics[width= 0.9\linewidth]{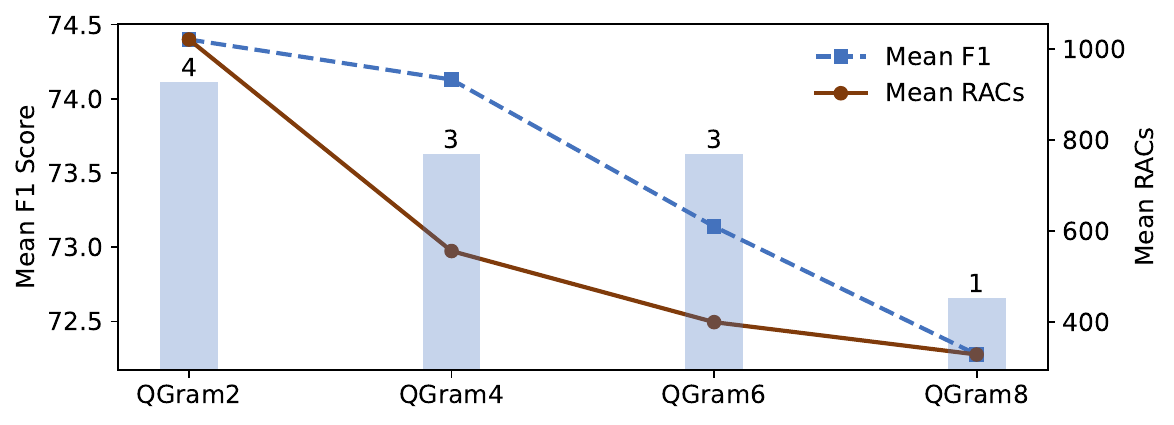}
  \caption{Batch Size Sensitivity. Effectiveness and efficiency vary with max batch size. Dashed line: mean F1 (left y-axis). Solid line: mean retrieval cost (right y-axis). Numbers above bars count datasets with the best F1.}
  \label{fig:blocksize}
\end{figure}

\noindent \textbf{Results and Analysis.}
Figure~\ref{fig:blocksize} summarizes the effectiveness--efficiency trade-off as we vary $b_{\max}$ with \texttt{QGram} blocking.
Overall, increasing $b_{\max}$ reduces retrieval overhead monotonically, as larger batches enable more reuse of a single retrieval result across multiple pairs (fewer \texttt{RACs}). In contrast, matching quality degrades gradually as $b_{\max}$ grows. This reflects a clear trade-off: larger batches improve efficiency by reducing retrieval calls, but they can dilute query specificity and yield less targeted context for some pairs, which may slightly hurt effectiveness. 
Among the tested configurations, $b_{\max} \in \{4,6\}$ provides a favorable balance, maintaining near-peak average \texttt{F1} while substantially reducing \texttt{RACs}.
We therefore use $b_{\max}=6$ as the default, prioritizing retrieval cost reduction while accepting a small decrease in \texttt{F1} relative to $b_{\max}=4$.

We further examine how batch size affects evidence quality and downstream matching performance on \texttt{FOZA}, \texttt{ITAM}, and \texttt{BEER}. Consistent with the mechanism characterized in Theorem~\ref{thm:acc_bs}, larger batches tend to reduce pair-specific evidence coverage and increase retrieval noise under a fixed context budget. 
For example, on \texttt{ITAM}, increasing the batch size from 1 to 8 raises coverage degradation by 0.27 and noise increase by 0.28; over the same range, recall decreases from 66.67\% to 48.15\%, while precision remains comparatively high. This pattern suggests that the resulting matching quality loss is primarily recall-driven. 
We further estimate the empirical coverage-degradation and noise-increase envelopes across batch sizes and show that their trends are consistent with the assumptions used in our theoretical analysis. The complete results can be found in Appendix~\ref{app:batch_theory_validation}. These results provide empirical support for the proposed evidence-quality mechanism and help explain why moderate batch sizes achieve a more favorable effectiveness–efficiency trade-off.

\section{Recommendations for Design Choices}
\label{recommendataion-design-choices}

Based on our empirical evaluation in \S\ref{sec:experiments}, we outline the following design recommendations for building a cost-effective \textit{RAG4EM}.

\noindent \textbf{Adopt Context-Adaptive Retrieval Granularity.}
As shown in \S\ref{subsec:C:retrieval_granularity}, node-level retrieval (QID/PID) is a lightweight and competitive default, while triple-level contexts become more beneficial as attribute diversity and ambiguity increase. We therefore recommend selectively shifting to EXP/BFS for harder cases, with EXP offering a favorable effectiveness--efficiency trade-off.

\noindent \textbf{Prioritize Batch Retrieval with Robust Blocking.}
The batching and blocking analyses show that batch retrieval reduces retrieval overhead while preserving per-pair inference~\S\ref{subsec:B}, and that $q$-gram-based blocking remains robust across datasets~\S\ref{subsec:C:blocking_strategy}. We therefore recommend batch retrieval with \texttt{QGram} as the default configuration.

\noindent \textbf{Use Batch Generation for Cost-Sensitive Settings.}
As shown in \S\ref{subsec:B}, batch generation substantially reduces inference latency and output-token cost, but can alter the precision--recall balance due to cross-pair coupling. It is therefore more suitable when inference cost or latency is the primary deployment constraint.

\noindent \textbf{Use Moderate Batch Sizes.}
The sensitivity analysis in \S\ref{subsec:D} shows that larger batches reduce retrieval cost but can weaken pair-specific evidence and degrade matching quality. In our evaluated datasets, $b_{\max}\in\{4,6\}$ provides a favorable effectiveness--efficiency trade-off.

Overall, a practical default is \texttt{QGram} blocking with batch retrieval, moderate batch size, and node-level knowledge, while introducing triple-level retrieval or batch generation when ambiguity or deployment cost becomes dominant.

\section{Related Work}
\label{sec:related}

Our work lies at the intersection of language model-based entity matching, blocking, and retrieval-augmented knowledge grounding. We review prior work along these three directions and highlight how \textit{CE-RAG4EM} differs from existing approaches.

\spara{PLM and LLM-based Entity Matching.}
Transformer-based pre-trained language models (PLMs), including BERT~\cite{LiDeepEM20, PeetersB21, TuFTWL0JG23, PaganelliTG24}, RoBERTa~\cite{TuFTWL0JG23, DingDWMZ24}, DistilBERT~\cite{PardoLGNNMBS25}, and GPT-2~\cite{ZhangAM2025}, have been widely applied to entity matching. These methods typically formulate EM as supervised sequence-pair classification and leverage contextualized representations to capture lexical and semantic correspondences between records. Their effectiveness has been further improved through task-specific fine-tuning~\cite{PeetersB21, TuFTWL0JG23}, knowledge injection~\cite{LiDeepEM20}, knowledge distillation~\cite{DingDWMZ24}, and graph-based contextualization~\cite{PardoLGNNMBS25}. Representative systems include GraLMatch~\cite{PardoLGNNMBS25}, which incorporates transitivity and graph context for group matching, and SETEM~\cite{DingDWMZ24}, which combines knowledge transfer and self-ensembling to improve PLM efficiency. Despite strong effectiveness, these approaches generally rely on labeled entity pairs for task-specific adaptation, which limits their applicability in domains where annotations are scarce or expensive to obtain.

Recent work increasingly explores LLM-based EM through in-context learning, serialization, and fine-tuning. SerializeEM~\cite{inKMSB25} introduces graph-structured serialization to expose richer relational information to the LLM. COMEM~\cite{WangCLCHSWZ25} studies multiple prompting paradigms, including matcher-, comparator-, and selector-based formulations, and combines them with ranking-based filtering for more robust matching. Mistral4SelectEM~\cite{RuanSB25} further improves selective matching through contrastive fine-tuning, aiming to distinguish true matches from hard negatives. Recent work has also explored batching- and clustering-based techniques to reduce LLM inference cost through in-context processing and grouped inference~\cite{fan2024cost, fu2025context}. These approaches demonstrate the flexibility of LLMs for EM, but they primarily optimize prompting, representation, fine-tuning, or inference organization. They do not explicitly study how external factual knowledge can be retrieved and shared across related candidate pairs in a zero-shot setting. In contrast, \textit{CE-RAG4EM} combines external knowledge grounding with the intermediate grouping structure already produced by EM blocking, allowing retrieval and generation overhead to be amortized across related pairs without introducing a separate batching stage.

\spara{Blocking for Entity Matching.}
Blocking is a standard component of EM pipelines that reduces the comparison space by grouping records sharing blocking keys or similarity signals~\cite{LiDeepEM20, TuFTWL0JG23, Nikoletos0K22}. Existing approaches span rule-based, token-based, clustering-based, graph-based, kNN-based, and learned blocking~\cite{Thirumuruganathan21, MoslemiBM24, WangZ24, GraphBlokcMu2023, ZhangWSDFP20, PaulsenGD23, WuWDHZ23, ZeakisPSK25}. Traditional techniques include standard blocking, sorted-neighborhood, QGram, and suffix blocking~\cite{AzzaliniJRT21}, while learned approaches such as DeepBlocker~\cite{Thirumuruganathan21} and meta-blocking~\cite{GagliardelliPSB22} introduce additional modeling and computation overhead. Prior work primarily uses blocking to prune unlikely matches and reduce pairwise comparisons. \textit{CE-RAG4EM} instead assigns blocking an additional role: it reuses the intermediate groups produced by blocking as batching units for retrieval and LLM inference. This allows related candidate pairs to share retrieved context and avoids introducing a separate clustering or batching stage.

\spara{Knowledge Grounding with RAG.}
Retrieval-augmented generation (RAG) grounds LLM outputs in externally retrieved context~\cite{LewisPPPKGKLYR020, GlassRCG21, jin2025longcontext}, while GraphRAG~\cite{HuLZPLZ25, GuoEmpoweringGraphRAG2025, ZhangGraphRAG4CustomiedLLM2025} and KG-RAG~\cite{MaKGRAG4SM25, 0002F0LMWY25} extend this paradigm with graph-structured knowledge. These approaches have been studied extensively in knowledge-intensive tasks such as question answering~\cite{ma2025llmkg4qa} and recommendation~\cite{PengGraphRAGSurvey2024}. 

Applying them to EM, however, introduces a different workload structure: EM involves many related pairwise decisions whose entities, attributes, and retrieval needs often overlap. A straightforward per-query RAG or KG-RAG pipeline therefore repeatedly retrieves and constructs similar context, while graph-based variants additionally incur traversal overhead. Existing work has not systematically studied this interaction between grounding quality and retrieval cost in LLM-based EM. \textit{CE-RAG4EM} addresses this gap by adapting textual and graph-based retrieval to the EM workload and amortizing retrieval over blocking-derived batches.

\section{Conclusions}
\label{sec:conclusion}
This paper proposes \textit{CE-RAG4EM}, a cost-efficient RAG framework with blocking-based batch processing to address the challenge of the \emph{grounding gap} in LLM-based entity matching and the costly retrieval and inference overhead in directly adapting vanilla RAG to entity matching in large-scale data integration settings.
Additionally, a design space for RAG-based entity matching was explored to systematically analyze and evaluate RAG4EM, focusing on blocking-aware batch optimizations and retrieval granularity. 
Extensive experiments across nine benchmark datasets demonstrate that \textit{CE-RAG4EM} achieves comparable or improved matching quality while substantially reducing end-to-end computation overhead and monetary cost, which well addresses the grounding gap in LLM-based entity matching.
Our systematic analysis reveals performance and overhead trade-offs that guide practitioners in designing scalable RAG systems for entity matching and data integration.

In the future, several directions that emerge from our findings can be further explored.
(i) \emph{Adaptive block size selection}: a promising direction is to predict the appropriate block size dynamically by estimating intra-block query similarity at the blocking phase. 
(ii) \emph{Cross-pair coupling mitigation}: batch generation improves efficiency but can degrade precision due to cross-pair coupling, while a prompt decomposition or output calibration could preserve the savings while enforcing per-pair independence.
(iii) \emph{Context-adaptive retrieval granularity}: an adaptive strategy that selects node- vs.\ triple-level retrieval based on attribute type could reduce overhead without sacrificing coverage. 
(iv) \emph{Integration with advanced blocking}: replacing unsupervised blocking with embedding-based methods could improve block purity and further amplify cost savings.

\begin{acks}
Chuangtao Ma and Arijit Khan acknowledge support from the Novo Nordisk Foundation grant NNF22OC0072415. This work was also supported by the GOBLIN COST Action (CA23147) and University of Amsterdam Data Science Centre.
We would like to thank the Wikidata Embedding Project team for providing the Wikidata VectorDB and its API access.
\end{acks}

\bibliographystyle{ACM-Reference-Format}
\balance
\bibliography{sample}

@inproceedings{PanahiWDN17,
  author       = {Fatemah Panahi and
                  Wentao Wu and
                  AnHai Doan and
                  Jeffrey F. Naughton},
  title        = {Towards Interactive Debugging of Rule-based Entity Matching},
  booktitle    = {{EDBT}},
  pages        = {354--365},
  year         = {2017}
}

@article{SinghMEMPQST17,
  author       = {Rohit Singh and
                  Venkata Vamsikrishna Meduri and
                  Ahmed K. Elmagarmid and
                  Samuel Madden and
                  Paolo Papotti and
                  Jorge{-}Arnulfo Quian{\'{e}}{-}Ruiz and
                  Armando Solar{-}Lezama and
                  Nan Tang},
  title        = {Synthesizing Entity Matching Rules by Examples},
  journal      = {Proc. {VLDB} Endow.},
  volume       = {11},
  number       = {2},
  pages        = {189--202},
  year         = {2017}
}

@inproceedings{MudgalLRDPKDAR18,
  author       = {Sidharth Mudgal and
                  Han Li and
                  Theodoros Rekatsinas and
                  AnHai Doan and
                  Youngchoon Park and
                  Ganesh Krishnan and
                  Rohit Deep and
                  Esteban Arcaute and
                  Vijay Raghavendra},
  title        = {Deep Learning for Entity Matching: {A} Design Space Exploration},
  booktitle = {{SIGMOD}},
  pages        = {19--34},
  year         = {2018}
}

@article{Thirumuruganathan21,
  author       = {Saravanan Thirumuruganathan and
                  Han Li and
                  Nan Tang and
                  Mourad Ouzzani and
                  Yash Govind and
                  Derek Paulsen and
                  Glenn Fung and
                  AnHai Doan},
  title        = {Deep Learning for Blocking in Entity Matching: {A} Design Space Exploration},
  journal      = {Proc. {VLDB} Endow.},
  volume       = {14},
  number       = {11},
  pages        = {2459--2472},
  year         = {2021}
}

@article{TuFTWL0JG23,
  author       = {Jianhong Tu and
                  Ju Fan and
                  Nan Tang and
                  Peng Wang and
                  Guoliang Li and
                  Xiaoyong Du and
                  Xiaofeng Jia and
                  Song Gao},
  title        = {Unicorn: {A} Unified Multi-tasking Model for Supporting Matching Tasks
                  in Data Integration},
  journal      = {Proc. {ACM} Manag. Data},
  volume       = {1},
  number       = {1},
  pages        = {84:1--84:26},
  year         = {2023}
}

@article{LiDeepEM20,
  author       = {Yuliang Li and
                  Jinfeng Li and
                  Yoshihiko Suhara and
                  AnHai Doan and
                  Wang{-}Chiew Tan},
  title        = {Deep Entity Matching with Pre-Trained Language Models},
  journal      = {Proc. {VLDB} Endow.},
  volume       = {14},
  number       = {1},
  pages        = {50--60},
  year         = {2020}
}

@article{PeetersB21,
  author       = {Ralph Peeters and
                  Christian Bizer},
  title        = {Dual-Objective Fine-Tuning of {BERT} for Entity Matching},
  journal      = {Proc. {VLDB} Endow.},
  volume       = {14},
  number       = {10},
  pages        = {1913--1921},
  year         = {2021}
}

@article{LiLSDT23,
  author       = {Yuliang Li and
                  Jinfeng Li and
                  Yoshi Suhara and
                  AnHai Doan and
                  Wang{-}Chiew Tan},
  title        = {Effective entity matching with transformers},
  journal      = {{VLDB} J.},
  volume       = {32},
  number       = {6},
  pages        = {1215--1235},
  year         = {2023}
}

@article{DingDWMZ24,
  author       = {Huahua Ding and
                  Chaofan Dai and
                  Yahui Wu and
                  Wubin Ma and
                  Haohao Zhou},
  title        = {{SETEM:} Self-ensemble training with Pre-trained Language Models for
                  Entity Matching},
  journal      = {Knowl. Based Syst.},
  volume       = {293},
  pages        = {111708},
  year         = {2024}
}

@inproceedings{ZhangGCS25,
  author       = {Zeyu Zhang and
                  Paul Groth and
                  Iacer Calixto and
                  Sebastian Schelter},
  title        = {A Deep Dive Into Cross-Dataset Entity Matching with Large and Small
                  Language Models},
  booktitle    = {{EDBT}},
  pages        = {922--934},
  year         = {2025}
}

@inproceedings{PeetersSB25,
  author       = {Ralph Peeters and
                  Aaron Steiner and
                  Christian Bizer},
  title        = {Entity Matching using Large Language Models},
  booktitle    = {{EDBT}},
  pages        = {529--541},
  year         = {2025}
}

@article{PaganelliTG24,
  author       = {Matteo Paganelli and
                  Donato Tiano and
                  Francesco Guerra},
  title        = {A multi-facet analysis of BERT-based entity matching models},
  journal      = {{VLDB} J.},
  volume       = {33},
  number       = {4},
  pages        = {1039--1064},
  year         = {2024}
}

@inproceedings{ZhangAM2025,
  author       = {Zeyu Zhang and
                  Paul Groth and
                  Iacer Calixto and
                  Sebastian Schelter},
  title        = {AnyMatch - Efficient Zero-Shot Entity Matching with a Small Language
                  Model},
  booktitle      = {GOOD DATA@AAAI},
  pages        = {1--10},
  year         = {2025}
}

@inproceedings{WangCLCHSWZ25,
  author       = {Tianshu Wang and
                  Xiaoyang Chen and
                  Hongyu Lin and
                  Xuanang Chen and
                  Xianpei Han and
                  Le Sun and
                  Hao Wang and
                  Zhenyu Zeng},
  title        = {Match, Compare, or Select? An Investigation of Large Language Models
                  for Entity Matching},
  booktitle    = {{COLING}},
  pages        = {96--109},
  year         = {2025}
}

@article{RuanSB25,
  author       = {Qian Ruan and
                  Dachuan Shi and
                  Thomas Bauernhansl},
  title        = {Fine-tuning large language models with contrastive margin ranking
                  loss for selective entity matching in product data integration},
  journal      = {Adv. Eng. Informatics},
  volume       = {67},
  pages        = {103538},
  year         = {2025}
}

@inproceedings{PeetersB23,
  author       = {Ralph Peeters and
                  Christian Bizer},
  title        = {Using ChatGPT for Entity Matching},
  booktitle    = {{ADBIS} (Short Papers)},
  volume       = {1850},
  pages        = {221--230},
  year         = {2023}
}

@ARTICLE{PKPN14,
  author={Papadakis, George and Koutrika, Georgia and Palpanas, Themis and Nejdl, Wolfgang},
  journal={{IEEE} Trans. Knowl. Data Eng.}, 
  title={Meta-Blocking: Taking Entity Resolution to the Next Level}, 
  year={2014},
  volume={26},
  number={8},
  pages={1946-1960}
}

@inproceedings{PardoLGNNMBS25,
  author       = {Fernando de Meer Pardo and
                  Claude Lehmann and
                  Dennis Gehrig and
                  Andrea Nagy and
                  Stefano Nicoli and
                  Branka Hadji Misheva and
                  Martin Braschler and
                  Kurt Stockinger},
  title        = {GraLMatch: Matching Groups of Entities with Graphs and Language Models},
  booktitle    = {{EDBT}},
  pages        = {1--12},
  year         = {2025}
}

@inproceedings{inKMSB25,
  author       = {Haoteng Yin and
                  Jinha Kim and
                  Prashant Mathur and
                  Krishanu Sarker and
                  Vidit Bansal},
  title        = {How to Talk to Language Models: Serialization Strategies for Structured
                  Entity Matching},
  booktitle    = {{NAACL} (Findings)},
  pages        = {7836--7850},
  year         = {2025}
}

@article{BodensohnUnveil2025,
  author       = {Jan{-}Micha Bodensohn and
                  Ulf Brackmann and
                  Liane Vogel and
                  Anupam Sanghi and
                  Carsten Binnig},
  title        = {Unveiling Challenges for LLMs in Enterprise Data Engineering},
  journal      = {Proc. {VLDB} Endow.},
  volume       = {19},
  number       = {2},
  pages        = {196--209},
  year         = {2025}
}

@article{KLZZZ26,
author = {Khan, Arijit and Luo, Yuyu and Zhang, Wenjie and Zhou, Minqi and Zhou, Xiaofang},
title = {Retrieval-augmented Generation (RAG): What is There for Data Management Researchers? A discussion on research from a panel at LLM+Vector Data Workshop @ IEEE ICDE 2025},
year = {2026},
volume = {54},
number = {4},
journal = {SIGMOD Rec.},
pages = {33–42}
}

@InProceedings{ma2025llmkg4qa,
  title={Large Language Models Meet Knowledge Graphs for Question Answering: Synthesis and Opportunities},
  author={Ma, Chuangtao and Chen, Yongrui and Wu, Tianxing and Khan, Arijit and Wang, Haofen},
  booktitle={EMNLP},
  year={2025},
  pages={24578--24597},
}

@article{MaKGRAG4SM25,
  author       = {Chuangtao Ma and
                  Sriom Chakrabarti and
                  Arijit Khan and
                  B{\'{a}}lint Moln{\'{a}}r},
  title        = {Knowledge Graph-based Retrieval-Augmented Generation for Schema Matching},
  journal      = {ACM Trans. Knowl. Discov. Data.},
  year         = {2026}
}

@inproceedings{LewisPPPKGKLYR020,
  author       = {Patrick Lewis and
                  Ethan Perez and
                  Aleksandra Piktus and
                  Fabio Petroni and
                  Vladimir Karpukhin and
                  Naman Goyal and
                  Heinrich K{\"{u}}ttler and
                  Mike Lewis and
                  Wen{-}tau Yih and
                  Tim Rockt{\"{a}}schel and
                  Sebastian Riedel and
                  Douwe Kiela},
  title        = {Retrieval-Augmented Generation for Knowledge-Intensive {NLP} Tasks},
  booktitle    = {NeurIPS},
  pages        = {1--16},
  year         = {2020}
}

@inproceedings{HuLZPLZ25,
  author       = {Yuntong Hu and
                  Zhihan Lei and
                  Zheng Zhang and
                  Bo Pan and
                  Chen Ling and
                  Liang Zhao},
  title        = {{GRAG:} Graph Retrieval-Augmented Generation},
  booktitle    = {{NAACL} (Findings)},
  pages        = {4145--4157},
  year         = {2025}
}

@article{PengGraphRAGSurvey2024,
  author       = {Boci Peng and
                  Yun Zhu and
                  Yongchao Liu and
                  Xiaohe Bo and
                  Haizhou Shi and
                  Chuntao Hong and
                  Yan Zhang and
                  Siliang Tang},
  title        = {Graph Retrieval-Augmented Generation: {A} Survey},
  journal      = {ACM Trans. Inf. Syst.},
  volume       = {44},
  number       = {2},
  pages        = {1--52},
  year         = {2025}
}

@article{ZhangGraphRAG4CustomiedLLM2025,
  author       = {Qinggang Zhang and
                  Shengyuan Chen and
                  Yuanchen Bei and
                  Zheng Yuan and
                  Huachi Zhou and
                  Zijin Hong and
                  Junnan Dong and
                  Hao Chen and
                  Yi Chang and
                  Xiao Huang},
  title        = {A Survey of Graph Retrieval-Augmented Generation for Customized Large
                  Language Models},
  journal      = {CoRR},
  volume       = {abs/2501.13958},
  year         = {2025}
}

@inproceedings{GuoEmpoweringGraphRAG2025,
  author       = {Kai Guo and
                  Harry Shomer and
                  Shenglai Zeng and
                  Haoyu Han and
                  Yu Wang and
                  Jiliang Tang},
  title        = {Empowering GraphRAG with Knowledge Filtering and Integration},
  booktitle      = {EMNLP},
  pages        = {25439--25453},
  year         = {2025}
}

@inproceedings{0002F0LMWY25,
  author       = {Shijie Wang and
                  Wenqi Fan and
                  Yue Feng and
                  Shanru Lin and
                  Xinyu Ma and
                  Shuaiqiang Wang and
                  Dawei Yin},
  title        = {Knowledge Graph Retrieval-Augmented Generation for LLM-based Recommendation},
  booktitle    = {{ACL}},
  pages        = {27152--27168},
  year         = {2025}
}

@inproceedings{ZhuXLLH25,
  author       = {Xiangrong Zhu and
                  Yuexiang Xie and
                  Yi Liu and
                  Yaliang Li and
                  Wei Hu},
  title        = {Knowledge Graph-Guided Retrieval Augmented Generation},
  booktitle    = {{NAACL-HLT}},
  pages        = {8912--8924},
  year         = {2025}
}

@inproceedings{MoslemiBM24,
  author       = {Mohammad Hossein Moslemi and
                  Harini Balamurugan and
                  Mostafa Milani},
  title        = {Evaluating Blocking Biases in Entity Matching},
  booktitle    = {{IEEE} Big Data},
  pages        = {64--73},
  year         = {2024}
}

@article{AzzaliniJRT21,
  author       = {Fabio Azzalini and
                  Songle Jin and
                  Marco Renzi and
                  Letizia Tanca},
  title        = {Blocking Techniques for Entity Linkage: {A} Semantics-Based Approach},
  journal      = {Data Sci. Eng.},
  volume       = {6},
  number       = {1},
  pages        = {20--38},
  year         = {2021}
}

@inproceedings{WangZ24,
  author       = {Runhui Wang and
                  Yongfeng Zhang},
  title        = {Pre-trained Language Models for Entity Blocking: {A} Reproducibility
                  Study},
  booktitle    = {{NAACL-HLT}},
  pages        = {8720--8730},
  year         = {2024}
}

@inproceedings{Nikoletos0K22,
  author       = {Konstantinos Nikoletos and
                  George Papadakis and
                  Manolis Koubarakis},
  title        = {{pyJedAI}: a Lightsaber for Link Discovery},
  booktitle    = {{ISWC} (Posters/Demos/Industry)},
  series       = {{CEUR} Workshop Proceedings},
  volume       = {3254},
  year         = {2022}
}

@inproceedings{Difallah2025wikirag,
  title={{WikiRAG}: Revisiting Wikidata KGC Datasets with Community Updates and Retrieval-Augmented Generation},
  author={Difallah, Djellel},
  booktitle={KDD},
  pages={5391--5401},
  year={2025}
}

@article{CaoGLXZX25,
  author       = {Yukun Cao and
                  Zengyi Gao and
                  Zhiyang Li and
                  Xike Xie and
                  S. Kevin Zhou and
                  Jianliang Xu},
  title        = {{LEGO-GraphRAG}: Modularizing Graph-based Retrieval-Augmented Generation
                  for Design Space Exploration},
  journal      = {Proc. {VLDB} Endow.},
  volume       = {18},
  number       = {10},
  pages        = {3269--3283},
  year         = {2025}
}

@inproceedings{primpeli2019wdc,
  title={The {WDC} Training Dataset and Gold Standard for Large-Scale Product Matching},
  author={Primpeli, Anna and Peeters, Ralph and Bizer, Christian},
  booktitle={WWW Companion},
  pages={381--386},
  year={2019}
}

@article{doan2020magellan,
  title={Magellan: toward building ecosystems of entity matching solutions},
  author={Doan, AnHai and Konda, Pradap and Suganthan GC, Paul and Govind, Yash and Paulsen, Derek and Chandrasekhar, Kaushik and Martinkus, Philip and Christie, Matthew},
  journal={Commun. {ACM}},
  volume={63},
  number={8},
  pages={83--91},
  year={2020},
}

@inproceedings{WikidataVrandecicPK23,
  author       = {Denny Vrandecic and
                  Lydia Pintscher and
                  Markus Kr{\"{o}}tzsch},
  title        = {Wikidata: The Making Of},
  booktitle    = {WWW Companion},
  pages        = {615--624},
  year         = {2023}
}

@inproceedings{GlassRCG21,
  author       = {Michael R. Glass and
                  Gaetano Rossiello and
                  Md. Faisal Mahbub Chowdhury and
                  Alfio Gliozzo},
  title        = {Robust Retrieval Augmented Generation for Zero-shot Slot Filling},
  booktitle    = {EMNLP},
  pages        = {1939--1949},
  year         = {2021}
}

@inproceedings{SturuaJina2025,
author = {Sturua, Saba and Mohr, Isabelle and Kalim Akram, Mohammad and G\"{u}nther, Michael and Wang, Bo and Krimmel, Markus and Wang, Feng and Mastrapas, Georgios and Koukounas, Andreas and Wang, Nan and Xiao, Han},
title = {Jina Embeddings V3: Multilingual Text Encoder with Low-Rank Adaptations},
year = {2025},
booktitle = {ECIR},
pages = {123–129},
numpages = {7}
}

@article{qwen3technicalreport,
      title={Qwen3 Technical Report}, 
      author={Qwen Team},
      year={2025},
      volume={abs/2505.09388},
      journal= {CoRR}
}

@inproceedings{kwon2023efficient,
  title={Efficient Memory Management for Large Language Model Serving with PagedAttention},
  author={Woosuk Kwon and Zhuohan Li and Siyuan Zhuang and Ying Sheng and Lianmin Zheng and Cody Hao Yu and Joseph E. Gonzalez and Hao Zhang and Ion Stoica},
  booktitle={SOSP},
  pages={611–626},
  year={2023}
}

@article{papa2021block,
author = {Papadakis, George and Skoutas, Dimitrios and Thanos, Emmanouil and Palpanas, Themis},
title = {Blocking and Filtering Techniques for Entity Resolution: A Survey},
year = {2020},
volume = {53},
number = {2},
journal = {ACM Comput. Surv.},
pages   = {31:1--31:42},
numpages = {42}
}

@article{tran2025kg20c,
  author       = {Hung{-}Nghiep Tran and
                  Atsuhiro Takasu},
  title        = {{KG20C} {\&} {KG20C-QA:} Scholarly Knowledge Graph Benchmarks
                  for Link Prediction and Question Answering},
  journal      = {CoRR},
  volume       = {abs/2512.21799},
  year         = {2025}
}

@article{Harnad2023LLM,
  author  = {Stevan Harnad},
  title   = {Language Writ Large: {LLMs}, {ChatGPT}, Meaning, and Understanding},
  journal      = {Frontiers Artif. Intell.},
  volume       = {7},
  pages      = {1--25},
  year         = {2024}
}

@inproceedings{ShaikhGKGYJ2023,
  author    = {Omar Shaikh and
               Kristina Gligori\'{c} and
               Ashna Khetan and
               Matthias Gerstgrasser and
               Diyi Yang and
               Dan Jurafsky},
  title     = {Grounding Gaps in Language Model Generations},
  booktitle = {{NAACL-HLT}},
  pages      = {6279--6296},
  year      = {2024}
}

@article{GagliardelliPSB22,
  author       = {Luca Gagliardelli and
                  George Papadakis and
                  Giovanni Simonini and
                  Sonia Bergamaschi and
                  Themis Palpanas},
  title        = {Generalized Supervised Meta-blocking},
  journal      = {Proc. {VLDB} Endow.},
  volume       = {15},
  number       = {9},
  pages        = {1902--1910},
  year         = {2022}
}

@article{GraphBlokcMu2023,
author = {Mugeni, John Bosco and Amagasa, Toshiyuki},
title = {A Graph-Based Blocking Approach for Entity Matching Using Contrastively Learned Embeddings},
year = {2023},
volume = {22},
number = {4},
journal = {SIGAPP Appl. Comput. Rev.},
pages = {37–46},
numpages = {10}
}

@inproceedings{ZhangWSDFP20,
  author       = {Wei Zhang and
                  Hao Wei and
                  Bunyamin Sisman and
                  Xin Luna Dong and
                  Christos Faloutsos and
                  David Page},
  title        = {AutoBlock: {A} Hands-off Blocking Framework for Entity Matching},
  booktitle    = {WSDM},
  pages        = {744--752},
  year         = {2020}
}

@article{PaulsenGD23,
  author       = {Derek Paulsen and
                  Yash Govind and
                  AnHai Doan},
  title        = {Sparkly: {A} Simple yet Surprisingly Strong {TF/IDF} Blocker for Entity
                  Matching},
  journal      = {Proc. {VLDB} Endow.},
  volume       = {16},
  number       = {6},
  pages        = {1507--1519},
  year         = {2023}
}

@article{WuWDHZ23,
  author       = {Shiwen Wu and
                  Qiyu Wu and
                  Honghua Dong and
                  Wen Hua and
                  Xiaofang Zhou},
  title        = {Blocker and Matcher Can Mutually Benefit: {A} Co-Learning Framework
                  for Low-Resource Entity Resolution},
  journal      = {Proc. {VLDB} Endow.},
  volume       = {17},
  number       = {3},
  pages        = {292--304},
  year         = {2023}
}

@article{ZeakisPSK25,
  author       = {Alexandros Zeakis and
                  George Papadakis and
                  Dimitrios Skoutas and
                  Manolis Koubarakis},
  title        = {An in-depth analysis of pre-trained embeddings for entity resolution},
  journal      = {{VLDB} J.},
  volume       = {34},
  number       = {1},
  pages        = {1--27},
  year         = {2025}
}

@inproceedings{10598022,
  author       = {George Papadakis and
                  Nishadi Kirielle and
                  Peter Christen and
                  Themis Palpanas},
  title        = {A Critical Re-evaluation of Record Linkage Benchmarks for Learning-Based
                  Matching Algorithms},
  booktitle    = {{ICDE}},
  pages        = {3435--3448},
  year         = {2024}
}

@inproceedings{fan2024cost,
  title={Cost-effective in-context learning for entity resolution: A design space exploration},
  author={Fan, Meihao and Han, Xiaoyue and Fan, Ju and Chai, Chengliang and Tang, Nan and Li, Guoliang and Du, Xiaoyong},
  booktitle={{ICDE}},
  pages={3696--3709},
  year={2024}
}

@article{fu2025context,
  title={In-context clustering-based entity resolution with large language models: A design space exploration},
  author={Fu, Jiajie and Tang, Haitong and Khan, Arijit and Mehrotra, Sharad and Ke, Xiangyu and Gao, Yunjun},
  journal={Proc. {ACM} Manag. Data},
  volume={3},
  number={4},
  pages={252:1--252:28},
  year={2025}
}

@inproceedings{jin2025longcontext,
title={Long-Context {LLM}s Meet {RAG}: Overcoming Challenges for Long Inputs in {RAG}},
author= {Bowen Jin and Jinsung Yoon and Jiawei Han and Sercan {\"{O}}. Arik},
booktitle={{ICLR}},
year={2025},
pages={1--39}
}

@article{chen2026generalized,
  title={Generalized Entity Matching with Adaptivity via Large Language Models},
  author={Chen, Xingguang and Shi, Yimin and Xiao, Xiaokui},
  journal={Proc. {ACM} Manag. Data},
  volume={4},
  number={3},
  pages={189:1--189:25},
  year={2026}
}

@inproceedings{pan2026understanding,
  author       = {Zhixuan Pan and
                  Shaowen Wang and
                  Pengfei Liao and
                  Jian Li},
  title        = {Understanding {LLM} Behaviors via Compression: Data Generation, Knowledge
                  Acquisition and Scaling Laws},
  booktitle    = {NeurIPS},
  pages        = {1--51},
  year         = {2025}
}

@inproceedings{roberts2020much,
  title = {How Much Knowledge Can You Pack Into the Parameters of a Language Model?},
  author={Roberts, Adam and Raffel, Colin and Shazeer, Noam},
  booktitle = {{EMNLP}},
  pages={5418--5426},
  year={2020}
}

\clearpage
\appendix
\nobalance

\section*{Appendix}    

This appendix provides prompt templates (Appendix~\ref{app:prompt}), proofs (Appendix~\ref{app:proof}), implementation details (Appendix~\ref{app:impl}), KG-RAG design choices (Appendix~\ref{app:expb}), additional experiments (Appendix~\ref{app:exp_addtional_top_k_context}--\ref{app:subsect_difficulty_levels}),  and qualitative analysis with examples and failure analysis (Appendix~\ref{app:qualitative_evaluation_analysis}).

\section{Prompt Templates}
\label{app:prompt}
\subsection{Prompt for Per-Query Inference}
\begin{tcolorbox}[boxrule=0.8pt, breakable]
    \noindent\textbf{Prompt for Per-Query Inference:} 
    
    You are an expert in entity matching, who is to determine whether these two given entity representations refer to the same entity. You are also provided with additional information retrieved from Wikidata, which might be helpful for your reasoning.

    \#\# \textbf{Input:} Entity 1: \{ \}  
    Entity 2: \{ \} 
    Additional Information (You can use this in your reasoning if available): \{ \}
                    
    \#\# \textbf{Instruction:} 1. Analyse each entity's semantics independently: consider key terms, roles, and context. 
    2. Rank the relevance of each entry in the additional information, and only use it if it helps make the decision. 
    3. Perform a step-by-step logical comparison of the two entities.

    \#\# \textbf{Output Format:} Match Decision: [Yes / No]
\end{tcolorbox}
\subsection{Prompt for Blocking-based Batch Inference}
\begin{tcolorbox}[boxrule=0.8pt, breakable]
\noindent\textbf{Prompt for Blocking-Based Batch Inference:}

    You are an expert in entity matching, who is to determine whether these two given entity representations refer to the same entity. You are also provided with additional information retrieved from Wikidata, which might be helpful for your reasoning.
    
    \#\# \textbf{Input}:
    Entity Pairs in a Batch:
    [Pair 1 - Entity 1: \{ \} Entity 2: \{ \}
    \ldots
    Pair N - Entity 1: \{ \} Entity 2: \{ \}]
    Additional Information (shared; you may use this in your reasoning if available): \{ \}
    
    \#\# \textbf{Instruction}:
      1. Process each entity pair sequentially, and treat each pair independently.
      2. Analyse each entity's semantics independently: consider key terms, roles, and context. 
      3. Rank the relevance of each entry in the additional information, and only use it if it helps make the decision. 
      4. Perform a step-by-step logical comparison of the two entities.
    
    \#\# \textbf{Output format:} Match Decisions: [Yes / No]
\end{tcolorbox}

\subsection{Prompt for LLM-as-Judge on Evidence Relevance} \label{app:prompt-llm-judge}

\begin{tcolorbox}[boxrule=0.8pt, breakable]
\noindent\textbf{Prompt for LLM-as-Judge on Evidence Relevance:}

    Determine whether this evidence is useful for deciding if the two records refer to the same real-world entity. Answer only Relevant or Irrelevant.

    \#\# \textbf{Input}:
    Entity 1: \{ \}
    Entity 2: \{ \}
    Retrieved Evidence: \{ \}

    \#\# \textbf{Output format}:
    [Relevant / Irrelevant]
\end{tcolorbox}

\section{Detailed Proof} \label{app:proof}

\subsection{Proof of Proposition 1}
\begin{proof}[Proof] 
The bounds follow directly from processing $k$ retrieved seed nodes per batch, considering $O(k^2)$ seed pairs for BFS, and executing approximately $\lceil N/\bar{b}\rceil$ batches.
\end{proof}

\subsection{Proof of Theorem 1}

\begin{defn}[Pair-level relevant evidence]
\label{def:relevant_evidence}
For a candidate pair $p=(r_1,r_2)$, let $R_p^\star \subseteq \mathcal{T}$ denote the set of evidence items relevant to the matching decision for $p$, where $\mathcal{T}$ is the universe of available evidence items in the knowledge base.
\end{defn}

\begin{defn}[Coverage and noise]
\label{def:cov_noise}
For a pair $p$ in batch $B$ with shared evidence context $C_B$, define
\[
\mathrm{cov}(p, B) = \frac{|C_B \cap R_p^\star|}{|R_p^\star|},
\qquad
\mathrm{noise}(p, B) = \frac{|C_B \setminus R_p^\star|}{|C_B|}.
\]
Coverage measures the fraction of relevant evidence retained in the shared context, whereas noise measures the fraction of retrieved evidence that is irrelevant to $p$. The corresponding per-query quantities are denoted by $\mathrm{cov}(p,\{p\})$ and $\mathrm{noise}(p,\{p\})$.
\end{defn}

\begin{defn}[Batch-induced evidence degradation]
\label{def:evidence_degradation}
For a batch $B$ of size $b$, define the average coverage degradation as
\[
\Delta_{\mathrm{cov}}(B) = \frac{1}{b}
\sum_{p\in B}
\left[\mathrm{cov}(p, \{p\}) - \mathrm{cov}(p,B)\right]_+,
\]
and the average noise increase as 
\[
\Delta_{\mathrm{noise}}(B) = \frac{1}{b}
\sum_{p\in B}
\left[\mathrm{noise}(p,B) - \mathrm{noise}(p,\{p\})\right]_+,
\]
where $[x]_+=\max\{x,0\}$. The positive-part operator allows batch retrieval to occasionally improve the evidence for an individual pair.
\end{defn}

\begin{assumption}[Batch-size-dependent context degradation]
\label{ass:batch_degradation}
There exist non-decreasing functions
$\delta_{\mathrm{cov}},\delta_{\mathrm{noise}}:
\mathbb{N}{+}\rightarrow[0,1]
$ such that, for every batch $B$ of size $b$, $\Delta_{\mathrm{cov}}(B) \leq \delta_{\mathrm{cov}}(b)$ and $\Delta_{\mathrm{noise}}(B) \leq \delta_{\mathrm{noise}}(b)$. Moreover, $\delta_{\mathrm{cov}}(1)
= \delta_{\mathrm{noise}}(1) = 0$.
\end{assumption}

The monotonicity assumption concerns only the upper bounds: increasing the batch size may increase the worst-case degradation, but does not imply that every larger batch necessarily has worse coverage or higher noise.

\begin{assumption}[Monotone--Lipschitz evidence response]
\label{ass:lipschitz}
There exist constants $L, L' \geq 0$ such that, for every candidate pair, the expected error rate $\mathrm{err}(\cdot)$ as a function of coverage $c$ and noise $n$ is non-increasing in $c$, non-decreasing in $n$, and jointly Lipschitz:
\[
\bigl| \mathrm{err}(c_1, n_1) - \mathrm{err}(c_2, n_2) \bigr| \leq L|c_1 - c_2| + L'|n_1 - n_2|.
\]
\end{assumption}

This assumption reflects the practical behavior of evidence-grounded LLM matching: higher coverage of relevant evidence generally improves matching reliability, whereas additional irrelevant evidence may increase ambiguity. The Lipschitz condition further assumes that the resulting change in prediction error is bounded with respect to changes in evidence quality. 

\begin{lemma}[Pair-level quality degradation]
\label{lem:pair_quality_degradation}
Under Assumption~\ref{ass:lipschitz}, for every pair $p\in B$,
\[
\begin{aligned}
&
\mathrm{err}\bigl(\mathrm{cov}(p,B), \mathrm{noise}(p,B)\bigr) - \mathrm{err}\bigl(\mathrm{cov}(p,\{p\}),
\mathrm{noise}(p,\{p\})\bigr)\\
&\qquad\leq
L\left[
\mathrm{cov}(p,\{p\}) - \mathrm{cov}(p,B)
\right]_+\\
&\qquad+
L'\left[
\mathrm{noise}(p,B) - \mathrm{noise}(p,\{p\})
\right]_+.
\end{aligned}
\]
\end{lemma}

\begin{proof}
Let $c_{\mathrm{pq}}=\mathrm{cov}(p, \{p\})$, $n_{\mathrm{pq}} = \mathrm{noise}(p, \{p\})$, $c_B=\mathrm{cov}(p,B)$, and $n_B=\mathrm{noise}(p,B)$. Define $\tilde{c}=\min\{c_{\mathrm{pq}},c_B\}$ and $\tilde{n}=\max\{n_{\mathrm{pq}},n_B\}$. By monotonicity, $\mathrm{err}(c_B,n_B)
\leq
\mathrm{err}(\tilde{c},\tilde{n})$. Applying the Lipschitz condition between $(\tilde{c},\tilde{n})$ and
$(c_{\mathrm{pq}},n_{\mathrm{pq}})$ gives
\[
\mathrm{err}(\tilde{c},\tilde{n}) - \mathrm{err}(c_{\mathrm{pq}},n_{\mathrm{pq}})
\leq
L|c_{\mathrm{pq}}-\tilde{c}| + L'|\tilde{n}-n_{\mathrm{pq}}|.
\]
By construction, $|c_{\mathrm{pq}}-\tilde{c}| = [c_{\mathrm{pq}}-c_B]_+$ and $|\tilde{n}-n_{\mathrm{pq}}| = [n_B-n_{\mathrm{pq}}]_+.$ Combining the inequalities proves the result.
\end{proof}

\newtheorem*{theorem*}{Theorem}
\begin{theorem*}[Batch-size-dependent error bound]
Under Assumptions~\ref{ass:batch_degradation} and~\ref{ass:lipschitz}, the average expected prediction-error increase of batch retrieval relative to per-query retrieval is bounded by
\[
\begin{aligned}
&
\frac{1}{b}
\sum_{p\in B} \Bigl[\mathrm{err}\bigl(\mathrm{cov}(p,B), \mathrm{noise}(p,B) \bigr) - \mathrm{err}\bigl(\mathrm{cov}(p,\{p\}), \mathrm{noise}(p,\{p\})\bigr)\Bigr]\\
&\qquad\leq
L\delta_{\mathrm{cov}}(b) + L'\delta_{\mathrm{noise}}(b) =:g(b).
\end{aligned}
\]
Moreover, $g(b)$ is non-decreasing in $b$ and satisfies $g(1)=0$.
\end{theorem*}

\begin{proof}
Averaging Lemma~\ref{lem:pair_quality_degradation} over all $p\in B$ yields
\[
\frac{1}{b}\sum_{p\in B}(\cdot) \leq L\Delta_{\mathrm{cov}}(B) + L’\Delta_{\mathrm{noise}}(B).
\]
Applying Assumption~\ref{ass:batch_degradation} gives the stated bound. Since $\delta_{\mathrm{cov}}(b)$ and $\delta_{\mathrm{noise}}(b)$ are non-decreasing and $L,L’\geq0$, $g(b)$ is also non-decreasing, with $g(1)=0$.
\end{proof}

We empirically examine the evidence-degradation mechanism underlying this theorem across different batch sizes; detailed evaluation protocol and results are provided in Appendix~\ref{app:batch_theory_validation}.

\section{Implementation Details}
\label{app:impl}
The details of the system implementation and configuration are given as follows.

\subsection{Blocking} The \texttt{QGram} blocking method is primarily integrated in the implementation of our \textit{CE-RAG4EM} via \texttt{pyJedAI}~\cite{Nikoletos0K22}. Additionally, the other blocking methods, i.e., Standard Blocking (\texttt{StdBlck}), Extended QGram \texttt{(XQGram)}, are evaluated. 
The parameters of blocking are set with the default configuration in \texttt{pyJedAI}: (i) the default size of $q$ in \texttt{QGram} and \texttt{(XQGram)} is 3; (ii) the threshold in the interval for controlling the number of combinations in \texttt{(XQGram)} is 0.9.
For a fair comparison, the hyperparameters across different blocking methods are kept the same as the default configuration.

\subsection{Vector Database}
\spara{Wikidata:} The Wikidata vector database\footnote{https://www.wikidata.org/wiki/Wikidata:Vector\_Database} and the API from the Wikidata embedding project\footnote{https://www.wikidata.org/wiki/Wikidata:Embedding\_Project} are utilized in the implementation of vector-based knowledge retrieval from Wikidata. 
The embeddings were trained by using a state‑of‑the‑art multilingual model, \texttt{jina‑embeddings‑v3}\footnote{https://huggingface.co/jinaai/jina-embeddings-v3}, with 512 dimensions, and then organized in DataStax Astra DB\footnote{https://www.ibm.com/products/datastax} for dense retrieval.
Specifically, DataStax Astra DB integrates JVector\footnote{https://github.com/datastax/jvector} search engine, where the approximate nearest neighbor (ANN) search is used for similarity search.
Additionally, the Wikidata REST API\footnote{https://www.wikidata.org/wiki/Wikidata:REST\_API} is called to fetch the correct label and search the connected items and properties to form the triples for the retrieved items or properties from Wikidata.

\spara{KG20C:} We trained the vector embeddings of entities in KG20C and organized them in the vector databases for dense retrieval, where the embedding model, vector database, and vector search engine are kept the same as the Wikidata database for fair comparison. 
Then, we employ triple search and expansion to locally search the neighborhood of triples connected to the entities that are identified from the initial dense search and ranking.

\subsection{Large Block Decomposition and Batch Size} \label{app:block_split_batch_size}

In the main evaluation, batches are derived from blocking groups and their decomposed sub-blocks, so the realized batch size varies across datasets, with a maximum size of \(6\). Any blocking group containing more than six candidate pairs is therefore split into non-overlapping sub-blocks of at most six pairs.

Additionally, we employ blocking to group all entities in the source table and target table instead of directly grouping candidate pairs. As a result, pairs within the same batch inherit shared blocking characteristics and often involve related records, providing a coherent basis for sharing retrieved context.
Table~\ref{tab:statistics-batch-size-spli-fraction} reports statistics on batch size and block-splitting fraction in CE-RAG4EM under varying maximum batch sizes with \textsc{QGram}.

\begin{table}[ht!]
\caption{Statistics on average batch sizes and block-splitting fraction (within parentheses) with \textsc{QGram} blocking.}
\label{tab:statistics-batch-size-spli-fraction}
\begin{center}
\small
\adjustbox{width=1\linewidth}{
\begin{tabular}{c|c|c|c|c}
\toprule
 & 2& 4& 6& 8\\ 
\midrule
 \textsf{ABT}& 1.91 (48.48\%)& 3.50 (34.85\%)& 4.18 (29.55\%)& 5.89 (26.52\%)\\
 \textsf{AMGO}& 1.88 (44.55\%)& 3.36 (31.19\%)& 4.49 (26.24\%)& 5.45 (20.30\%)\\
 \textsf{BEER}& 1.89 (66.67\%)& 3.37 (44.44\%)& 4.55 (44.44\%)& 5.35 (44.44\%)\\
 \textsf{DBAC}& 1.94 (48.62\%)& 3.66 (36.70\%)& 5.12 (31.19\%)& 6.49 (25.69\%)\\ 
 \textsf{DBGO}& 1.96 (50.60\%)& 3.78 (39.76\%)& 5.44 (36.14\%)& 6.97 (31.93\%)\\ 
 \textsf{FOZA} & 1.87 (50.00\%)& 3.32 (40.00\%)& 4.40 (35.00\%)& 5.25 (25.00\%)\\
 \textsf{ITAM} & 1.76 (37.50\%)& 2.73 (25.00\%)& 3.21 (25.00\%)& 3.52 (20.83\%)\\
 \textsf{WAAM} & 1.88 (45.89\%)& 3.29 (31.40\%)& 4.39 (25.12\%)& 5.19 (20.77\%)\\
 \textsf{WDC}& 1.88 (42.27\%)& 3.33 (33.82\%)& 4.40 (29.09\%)& 5.27 (23.27\%)\\
 \bottomrule
\end{tabular}}
\end{center}
\end{table}
As shown, increasing the maximum block size from 2 to 8 leads to a higher \textit{avg. batch size} and a lower \textit{block-splitting fraction}.
The choice of batch size was guided by LLM context length constraints and the trade-off between matching quality and cost considerations: larger batches reduce retrieval cost but may slightly affect performance. 
The batch size sensitivity analysis is given in \S~\ref{subsec:D}.

\section{KG-RAG Design Choices}
\label{app:expb}

\noindent\textbf{Research Question.}
\textit{Which graph traversal strategy and triple budget are more effective for KG-RAG-based entity matching?}

\noindent \textbf{Evaluation Protocol.}
We evaluate KG-RAG design choices by instantiating \textit{CE-KG-RAG4EM-BR} under the same batch-retrieval setting and varying only the triple construction procedure. For each block, we retrieve the top-ranked Wikidata items via vector search and use them as seeds to construct candidate triples using either expansion-based traversal (EXP) or breadth-first search (BFS). To control prompt length and reduce noise, we apply the same ranking-based subgraph refinement for both traversal strategies.
We further vary the retained triple budget, keeping the Top-$k$ refined triples with $k \in \{1,2\}$, resulting in four configurations: EXP Top-1, EXP Top-2, BFS Top-1, and BFS Top-2.

\begin{figure}[ht]
  \centering
  \includegraphics[width=1 \linewidth]{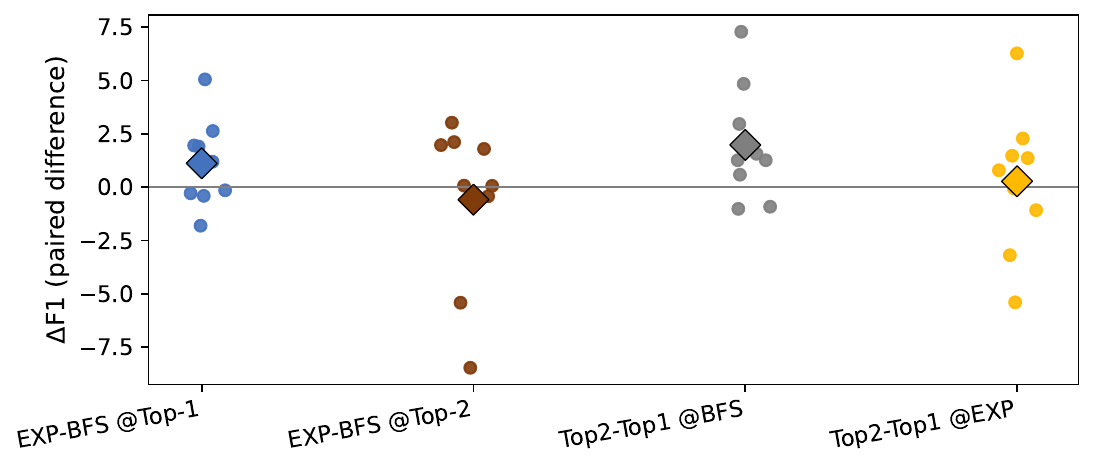}
  \caption{KG-RAG Design Choices. Paired $\Delta$F1 across datasets for traversal strategy (EXP$-$BFS at Top-1/Top-2) and triple budget (Top-2$-$Top-1 within BFS/EXP). Each dot is one dataset; diamonds show mean differences.}
  \label{fig:kg_rag_choice}
\end{figure}

\noindent \textbf{Results and Analysis.}
Figure~\ref{fig:kg_rag_choice} reports dataset-level paired \texttt{F1} differences for two KG-RAG design axes: (i) traversal strategy (EXP vs.\ BFS) under a fixed triple budget, and (ii) triple budget (Top-2 vs.\ Top-1) under a fixed traversal strategy. Each point denotes a dataset-level paired difference and the diamond indicates the mean.

\noindent \underline{Traversal strategy (EXP vs.\ BFS).}
At a fixed triple budget, EXP and BFS exhibit mixed wins across datasets, and neither traversal dominates consistently. EXP constructs a concise 1-hop neighborhood around the retrieved seed QIDs, which is particularly effective when matching relies on strong, near-exact identifiers and when the retrieved seeds are accurate. In contrast, BFS can incorporate multi-hop relations useful for semantically ambiguous records, though this broader traversal also increases the risk of off-topic relations.

\noindent \underline{Triple budget (Top-2 vs.\ Top-1).}
Increasing the triple budget from Top-1 to Top-2 tends to help BFS more than EXP. BFS benefits from retaining multiple complementary triples along different paths, whereas EXP prioritizes the most salient one-hop relations early, yielding diminishing returns from a second triple. Although increasing the BFS triple budget can make BFS competitive with EXP on many datasets, this improvement comes at a substantial expansion cost, consistent with the overhead trends (Figure~\ref{fig:rg-efficiency-b}).

\section{Additional Exploration of Top-$k$ Context}
\label{app:exp_addtional_top_k_context}

\noindent\textbf{Research Question.}
\textit{How does the number of the top-$k$ retrieved context impact the overall matching performance and cost?}

\noindent \textbf{Evaluation Protocol.} Our main evaluation uses a small number of retrieved contexts as evidence, since excessive contextual information can introduce noise and mislead LLM reasoning~\cite{jin2025longcontext}. 
To assess sensitivity to the retrieval budget, we extend the evaluation to explore the impact of the top-$k$ contexts ranging from $1$ to $5$ and evaluate the effects on matching performance and cost.

\begin{table}[ht]
\centering
\caption{Impact of the Top-$k$ \texttt{PID} on performance and cost.}
\label{tab:top_k_exploration}
\label{small}
\adjustbox{width=1\linewidth}{
\begin{tabular}{c|c|c|c|c|c|c}
\toprule
  Dataset & Metric & $k$=1&
  $k$=2&
  $k$=3& $k$=4&$k$=5\\
\midrule
\multirow{3}{*}{\textsf{FOZA}}& \texttt{F1} (\%)& \underline{79.98}& \textbf{83.11}&  76.62& 77.72&78.88\\
                      & \textsc{Token} (M)& \textbf{0.1557}& \underline{0.1600}&  0.1668& 0.1659&0.1708\\
                      & \textsc{ Time} (s)& \textbf{0.18}& \underline{0.20}&  0.26& 0.27&0.24\\ \hline
\multirow{3}{*}{\textsf{ITAM}}& \texttt{F1} (\%)& 59.84& \textbf{71.08}&  \underline{63.32}& \underline{63.32}&61.69\\
                      & \textsc{Token} (M)& \textbf{0.1108}& \underline{0.1128}&  0.1141& 0.1172&0.1180\\
                      & \textsc{Time} (s)& \textbf{0.20}& \underline{0.21}&  0.30& 0.39& 0.30\\ \hline
 \multirow{3}{*}{\textsf{BEER}}& \texttt{F1} (\%)& 61.99& \textbf{73.49}& \underline{68.65}& 65.56&64.65\\
 & \textsc{Token} (M)& \textbf{0.0718}& \underline{0.0727}& 0.0762& 0.0775&0.0778\\
 & \textsc{Time} (s)& \textbf{0.22}& \underline{0.23}& \underline{0.23}& 0.24&0.26\\
 \bottomrule
\end{tabular}}
\end{table}

\noindent \textbf{Results and Analysis.} Table~\ref{tab:top_k_exploration} reports the overall matching performance (\texttt{F1} score) and cost (total token consumption and average inference time) on \textsc{FOZA}, \textsc{ITAM}, and \textsf{BEER} datasets. 
The results demonstrate that increasing the context size $\ge$ top-2 leads to a decrease in overall performance and an increase in token consumption and inference time.
This is consistent with the previous findings that increasing the number of retrieved contextual knowledge may mislead LLMs and further leads to a performance drop in the RAG system~\cite{jin2025longcontext}.
These results also confirm that our design choice for the evaluation of the impact of the top-$k$ from 1 and 2 in the main evaluation was based on balancing overall performance and cost.

\section{Empirical Analysis of Batch Size, Evidence Quality, and Matching Performance}
\label{app:batch_theory_validation}

We empirically examine how batch size affects retrieval quality and matching performance on three representative datasets, thereby evaluating the evidence-degradation mechanism underlying Theorem~\ref{thm:acc_bs} and the trade-off between retrieval noise and batch size.

\noindent \textbf{Evaluation Protocol.}
Following the main experimental setting, we fix the final retrieved top-$2$ ranked evidence items (QID) for every batch and vary only the batch size. Since constructing pair-level relevance labels requires judging a large number of retrieved evidence items, we restrict this analysis to \textsf{BEER}, \textsf{FOZA}, and \textsf{ITAM} to keep the LLM-based evaluation cost manageable. 

Because the complete relevant-evidence set $R_p^\star$ is not directly observable, for each candidate pair $p$ we construct an empirical reference set $\widehat{R}_p$ from the union of evidence retrieved across all evaluated batch-size settings. Given the two records and one retrieved evidence item, we use the instruction in Appendix~\ref{app:prompt-llm-judge} and \texttt{gpt-5-mini} as an independent LLM judge to classify the evidence as \emph{Relevant} or \emph{Irrelevant}. 
Then all retrieved candidate items labeled as \emph{Relevant} are collected to form the evidence set $\widehat{R}_p$, which is subsequently passed to the downstream evaluation.
We then substitute $\widehat{R}p$ for $R_p^\star$ in Definitions~\ref{def:cov_noise}–\ref{def:evidence_degradation} to estimate coverage degradation $\Delta{\mathrm{cov}}$ and noise increase $\Delta_{\mathrm{noise}}$. To relate the observations to Assumption~\ref{ass:batch_degradation}, we define empirical degradation across datasets as
\[
\widehat{\delta}_{\mathrm{cov}}(b)
=
\max_D \Delta_{\mathrm{cov}}^{(D)}(b),
\qquad
\widehat{\delta}_{\mathrm{noise}}(b)
=
\max_D \Delta_{\mathrm{noise}}^{(D)}(b),
\]
where $D$ ranges over datasets, i.e., \textsf{BEER}, \textsf{FOZA}, and \textsf{ITAM}.

\begin{table}[ht]
\centering
\renewcommand{\arraystretch}{1}

\caption{Evidence-quality degradation and matching performance under different batch sizes with Top-2 QIDs.}
\label{tab:batch_theory}
\small
\adjustbox{width=1\linewidth}{
\begin{tabular}{c|c|ccccc}
\toprule
Dataset & Metric & $b=1$ & $b=2$ & $b=4$ & $b=6$ & $b=8$ \\
\midrule

\multirow{6}{*}{\textsf{FOZA}}
& $\Delta_{\mathrm{cov}}$
& 0.00 & 0.12 & 0.15 & 0.15 & 0.16 \\
& $\Delta_{\mathrm{noise}}$
& 0.00 & 0.10 & 0.12 & 0.11 & 0.13 \\
& Error (\%)
& 2.65 & 3.17 & 3.17 & 4.23 & 3.70 \\
& Precision (\%)
& 100.00 & 100.00 & 100.00 & 93.75 & 100.00 \\
& Recall (\%)
& 77.27 & 72.73 & 72.73 & 68.18 & 68.18 \\
& F1 (\%)
& 87.18 & 84.21 & 84.21 & 78.95 & 81.08 \\
\midrule

\multirow{6}{*}{\textsf{ITAM}}
& $\Delta_{\mathrm{cov}}$
& 0.00 & 0.18 & 0.23 & 0.25 & 0.27 \\
& $\Delta_{\mathrm{noise}}$
& 0.00 & 0.19 & 0.24 & 0.25 & 0.28 \\
& Error (\%)
& 8.26 & 10.09 & 9.17 & 11.93 & 13.76 \\
& Precision (\%)
& 100.00 & 100.00 & 100.00 & 93.75 & 92.86 \\
& Recall (\%)
& 66.67 & 59.26 & 62.96 & 55.56 & 48.15 \\
& F1 (\%)
& 80.00 & 74.42 & 77.27 & 69.77 & 63.41 \\
\midrule

\multirow{6}{*}{\textsf{BEER}}
& $\Delta_{\mathrm{cov}}$
& 0.00 & 0.07 & 0.09 & 0.09 & 0.11 \\
& $\Delta_{\mathrm{noise}}$
& 0.00 & 0.06 & 0.07 & 0.07 & 0.09 \\
& Error (\%)
& 6.59 & 7.69 & 7.69 & 7.69 & 8.79 \\
& Precision (\%)
& 90.00 & 100.00 & 88.89 & 88.89 & 87.50 \\
& Recall (\%)
& 64.29 & 50.00 & 57.14 & 57.14 & 50.00 \\
& F1 (\%)
& 75.00 & 66.67 & 69.57 & 69.57 & 63.64 \\
\midrule

& $\widehat{\delta}_{\mathrm{cov}}(b)$
& 0.00 & 0.18 & 0.23 & 0.25 & 0.27 \\
& $\widehat{\delta}_{\mathrm{noise}}(b)$
& 0.00 & 0.19 & 0.24 & 0.25 & 0.28 \\

\bottomrule
\end{tabular}}
\end{table}

\noindent \textbf{Results and Analysis.} Table~\ref{tab:batch_theory} reveals a clear trade-off between batch size, retrieval quality, and matching performance. As the batch size increases, the empirical degradation envelopes increase from $0$ to $0.27$ for coverage and from $0$ to $0.28$ for noise. Although individual datasets exhibit small fluctuations at intermediate batch sizes, the worst-case is non-decreasing, consistent with Assumption~\ref{ass:batch_degradation}.
The downstream effect is primarily recall-driven rather than a uniform loss of precision. On \textsf{ITAM}, for example, increasing $b$ from $1$ to $8$ raises $\Delta_{\mathrm{cov}}$ from $0$ to $0.27$ and $\Delta_{\mathrm{noise}}$ from $0$ to $0.28$. Over the same range, recall decreases from $66.67\%$ to $48.15\%$, while precision remains comparatively high, decreasing from $100.00\%$ to $92.86\%$. Consequently, F1 falls from $80.00\%$ to $63.41\%$ and the error rate increases from $8.26\%$ to $13.76\%$. A similar recall reduction is observed on \textsf{FOZA}, where recall decreases from $77.27\%$ to $68.18\%$, whereas precision remains high across batch sizes.

These observations indicate that larger shared contexts can affect matching through two related mechanisms: relevant pair-specific evidence may be diluted as coverage decreases, leading to missed true matches, while additional evidence retrieved for other pairs increases noise and can occasionally introduce false positives. In our evaluated datasets, the former effect is more pronounced, as the degradation in F1 is driven mainly by lower recall rather than a consistent decrease in precision. Overall, the results provide empirical evidence for the retrieval-noise/batch-size trade-off and support the evidence-quality mechanism underlying Theorem~\ref{thm:acc_bs}.

\section{Scalability Across KG Domains}
\label{app:expc}

\noindent\textbf{Research Question.}
\textit{Does \textit{CE-RAG4EM} maintain its effectiveness across different domains of knowledge graphs: commonsense versus domain-specific KGs?}

\noindent \textbf{Evaluation Protocol.} We examine the scalability of \textit{CE-RAG4EM} from commonsense KGs to domain-specific KGs with lower knowledge overlap between the LLM's internal knowledge and the retrieved external knowledge.
We evaluate and compare \textit{CE-KG-RAG4EM-BR} with Wikidata and KG20C on \textsf{DBAC} and \textsf{DBGO} datasets, using Top-$2$ refined EXP triples from the retrieved entities.

\noindent \textbf{Results and Analysis.}
Table~\ref{tab:scalability_kg_domain} compares the performance of LLM, \textit{CE-KG-RAG4EM-BR}, with the top-2 refined EXP triple from the commonsense knowledge graph (Wikidata) versus a domain-specific knowledge graph (KG20C). 
\textit{CE-RAG4EM} substantially outperforms LLM-based entity matching both in terms of \texttt{R} and \texttt{F1} on \textsf{DBAC} and \textsf{DBGO} both in KG20C and Wikidata, which demonstrate that our \textit{CE-RAG4EM} successfully maintains its effectiveness across knowledge graph domains.
Additionally, \textit{CE-RAG4EM} with triple from domain-specific KG (KG20C) slightly outperforms commonsense KG (Wikidata) in terms of \texttt{R} and \texttt{F1} on \textsf{DBAC}. 
On \texttt{DBGO}, KG20C attains slightly higher \texttt{P}, while Wikidata remains stronger in \texttt{R} and \texttt{F1}.
This result confirms that \textit{CE-RAG4EM} does not suffer performance degradation when extending KG from commonsense to domain-specific, which also highlights that \textit{CE-RAG4EM} with a knowledge graph where there is lower knowledge overlap, i.e, domain-specific KG, works better than commonsense KG on the domain-specific entity matching dataset.
However, we also observed that the contextual knowledge retrieved from the domain-specific KG is not always useful for domain-specific entity matching, which mainly depends on the relevance of the retrieved contextual knowledge. 
Additional details for these cases are provided in case study 2--3 in Appendix~\ref{app:addtional_examples}.

\begin{table}[ht]
\centering
\caption{Scalability Across KG Domains. KG20C vs.\ Wikidata of \textit{CE-KG-RAG4EM-BR}.}
\label{tab:scalability_kg_domain}
\small
\adjustbox{width=1\linewidth}{
\begin{tabular}{c|c|c|c|c}
\toprule
  Dataset & Metric (\%) & LLM &
  \makecell{KG20C\\ Top-2 Triple (EXP)} &
  \makecell{Wikidata\\ Top-2 Triple (EXP)} \\
\midrule
\multirow{3}{*}{\textsf{DBAC}} & Precision  & \textbf{96.15} & 94.80 & 94.91 \\
                      & Recall  & 63.44 & \textbf{72.60} & 71.32 \\
                      & F1 & 76.43 & \textbf{82.23} & 81.44 \\ 
\midrule                  
\multirow{3}{*}{\textsf{DBGO}} & Precision  & 93.13 & \textbf{93.63} & 92.62 \\
                      & Recall  & 49.98 & 65.92 & \textbf{67.70} \\
                      & F1 & 65.02 & 77.36 & \textbf{78.15} \\
\bottomrule
\end{tabular}}
\end{table}

\section{Evaluation on Subsets of Pairs with Different Difficulty Levels} \label{app:subsect_difficulty_levels}
To examine how retrieved context affects EM across different difficulty levels of pairs, we conduct a quantitative analysis.

\noindent \textbf{Evaluation Protocol.} 
We group record pairs into three difficulty subsets based on vector similarity. Specifically, we encode the serialized representation of each record using the \texttt{all-MiniLM-L6-v2} sentence encoder\footnote{https://huggingface.co/sentence-transformers/all-MiniLM-L6-v2} and compute the similarity between the two records in each pair. We then define a lower threshold at the $q_l$-th percentile and an upper threshold at the $q_h$-th percentile of the empirical similarity distribution, with $q_l$=0.33 and $q_h$=0.67. Using percentile-based thresholds provides dataset-relative cutoffs and yields approximately balanced subsets across datasets.

Each pair is assigned to one of three subsets: \textit{Easy Non-Match} (ENM), where the similarity score is at or below the lower threshold; \textit{Ambiguous} (AM), where the score lies between the two thresholds and the records exhibit intermediate similarity; and \textit{Easy Match} (EM), where the score is at or above the upper threshold. These subsets provide a similarity-based proxy for matching difficulty, with the ambiguous subset capturing pairs for which surface-level record similarity provides less decisive evidence.

We compare the base \textit{LLM-only} setting with \textit{CE-RAG4EM} using the Top-$2$ ranked PIDs as retrieved context. For each subset, we report the number of pairs in each subset (\# Pair), the changes in precision ($\Delta$\texttt{P}), recall ($\Delta$\texttt{R}), and F1 score ($\Delta$\texttt{F1}) relative to the LLM-only baseline, together with \textit{success rate} (\texttt{SR}) and \textit{failure rate} (\texttt{FR}). \texttt{SR} denotes the fraction of pairs incorrectly predicted by the LLM-only baseline but corrected by \textit{CE-RAG4EM}, whereas \texttt{FR} denotes the fraction of pairs correctly predicted by the baseline but incorrectly predicted after retrieval augmentation.

\begin{table}[ht]
\centering
\renewcommand{\arraystretch}{1}
\caption{Quantitative evaluation on a subset of pairs with different difficulty levels.}
\label{tab:evaluation_different_difficulty_levels}
\small
\adjustbox{width=1\linewidth}{
\begin{tabular}{c|c|c|c|c|c|c|c}
\toprule
  {Dataset}&Subset & \# Pair&$\Delta$\texttt{P} (\%)&$\Delta$\texttt{R} (\%)&$\Delta$\texttt{F1} (\%)&\texttt{SR} (\%)&\texttt{FR} (\%)\\
\midrule
\multirow{3}{*}{\textsf{AMGO}}& ENM& 757& 0&   99.93&77.76& 0.04&0.09\\
                      & AM& 779& -16.54&   2.32&-4.82& 1.07&1.54\\
                      & EM& 757& -10.76&   8.13&18.63& 7.84&7.62\\ 
\midrule
\multirow{3}{*}{\textsf{WAAM}}& ENM& 676& 3.23&   7.68&3.46& 0.10&0.30\\
                      & AM& 697& 6.69&   6.78&6.70& 0.81&1.00\\
                      & EM& 676& -2.20&   5.08&1.99& 1.19&1.34\\
\midrule
\multirow{3}{*}{\textsf{DBAC}}& ENM& 816& 0&  0&0& 0&0\\
 & AM& 841& 14.86&  62.51&45.56& 0.44&0.08\\
 & EM& 816& -1.18&  21.24&12.05& 10.95&3.64\\
 \bottomrule
\end{tabular}}
\end{table}

\noindent \textbf{Results and Analysis.}
Table~\ref{tab:evaluation_different_difficulty_levels} compares \textit{CE-RAG4EM} with LLM-EM on the difficulty-based pair subsets from \textsc{AMGO}, \textsc{WAAM}, and \textsc{DBAC}, where ambiguous records with sparse or mixed textual and numeric attributes are commonly observed.
The results reveal that retrieval has little effect on \textit{Easy Non-Match} pairs, for which the LLM-EM already performs reliably. Both \texttt{SR} and \texttt{FR} remain close to zero across the three datasets (e.g., both are 0\% on \textsf{DBAC}), indicating that retrieval rarely changes decisions for clearly dissimilar pairs.

For \textit{Easy Match} pairs, retrieval consistently improves \texttt{F1}, with \(\Delta\)\texttt{F1} of \(+18.63\%\), \(+1.99\%\), and \(+12.05\%\) across the three datasets. The highest observed success rate is \(10.95\%\) on \textsf{DBAC}. These results suggest that retrieved context can reinforce matching evidence for highly similar record pairs and help recover additional true matches by external knowledge grounding. Consistently, \(\Delta\)\texttt{R} remains non-negative across these subsets, indicating that retrieval helps the LLMs identify additional true matches without missing the previously detected ones. 

For \textit{Ambiguous} pairs, the benefit of retrieval is more dataset-dependent. On \textsf{DBAC}, \textit{CE-RAG4EM} improves \texttt{F1} by \(45.56\%\), with \texttt{SR}=0.44\% and \texttt{FR}=0.08\%, indicating that the retrieved context is largely beneficial. In contrast, on \textsf{AMGO}, \texttt{F1} decreases by \(4.82\%\), with \texttt{FR} exceeding \texttt{SR} (1.54\% vs.\ 1.07\%). This suggests that, for ambiguous pairs, weakly relevant or noisy retrieved evidence can sometimes outweigh its grounding benefit, consistent with the precision degradation observed in Table~\ref{tab:cerag4em_llmem_metrics}. A detailed failure example (\textbf{E1}) is provided in Appendix~\ref{appsub:failure-analysis}.

Overall, our analysis reveals that retrieval in \textit{CE-RAG4EM} is consistently beneficial for easy-match pairs, but shows limited benefit for easy non-match pairs.
The benefit of retrieval to ambiguous pairs depends on the quality and relevance of the retrieved context, confirming that retrieval is not universally advantageous and should be complemented by mechanisms to mitigate noisy context.

\section{Qualitative Evaluation and Analysis}
\label{app:qualitative_evaluation_analysis}

\subsection{Qualitative Analysis with Example} \label{app:case_study}
To show the hallucination mitigation of \textit{CE-RAG4EM} with the retrieved context, we conduct a qualitative analysis with an example. 
Specifically, we illustrate how \textit{CE-RAG4EM} resolves an ambiguous record pair from the \textsf{FOZA} restaurant dataset with varying types of attributes, including text, numeric, and categorical attributes. 
\begin{tcolorbox}[boxrule=0.8pt, colback=blue!5, breakable]
\noindent\textbf{Example 1: Hallucination Mitigation}

    \textbf{EM Pair}: \textit{Entity 1}: name: ` yujean kang's gourmet chinese cuisine '; addr: '67 n. raymond ave. '; city: ` los angeles '; phone: 818/585 -0855; type: asian; class: 22.  \textit{Entity 2}: name: ` yujean kang's '; addr: '67 n. raymond ave. '; city: pasadena; phone: 818-585-0855; type: chinese; class: 22

    \textbf{Ground Truth}: 1 (Yes, Matched)
    
    \textbf{Retrieved Triple:} (Chinese-Indonesian restaurant, subclass of, Chinese restaurant); (Chinese-Indonesian restaurant, instance of, Inventory Intangible Cultural Heritage in the Netherlands);(Pasadena, instance of, big city); (Pasadena, instance of, city of California); (Pasadena, instance of, charter city)
    
    \textbf{LLM Output}: \textcolor{red}{No}; Given that these are distinct cities, it is highly unlikely that they refer to the same entity, even though they share the same address and have similar names and phone numbers.
    
    \textbf{CE-KG-RAG4EM Output}: \textcolor{LimeGreen}{Yes}; \textit{The provided contextual knowledge is relevant for understanding the location and geographical context. The context of the cities being close and the type of cuisine being related further support this conclusion.} Despite the differences in city names (Los Angeles vs. Pasadena), the matching address, similar names, and consistent phone numbers suggest a strong likelihood that both entities refer to the same restaurant, possibly with a slight variation in naming or categorization. 
\end{tcolorbox}
As shown in the above example, \textit{CE-KG-RAG4EM} with the retrieved Top-2 ranked EXP triples from Wikidata outputs the correct answer in the ambiguous matching case. Specifically, the retrieved triples: \texttt{(Pasadena, instance of, city of California)}; \texttt{(Pasadena, instance of, charter city)} reveal that \texttt{Pasadena} is a charter city in California, which further provides a reasoning chain for the LLM to infer potential connections between \texttt{Pasadena} and \texttt{Los Angeles}. Namely, \texttt{Pasadena} is a charter city in \texttt{Los Angeles County, California}, which is helpful for LLM's decision-making, as it guides the model to pay attention to the connection between these two cities and generate the correct answer.

\subsection{Additional Analysis with Example} \label{app:addtional_examples}

We provide an additional qualitative analysis with examples to further demonstrate the effectiveness and scalability of extending \textit{CE-RAG4EM} from the commonsense KG (Wikidata) to the domain-specific KG (KG20C) on the \textsc{DBGO} dataset.
Specifically, \textbf{Example 2} shows that the triples retrieved from the domain-specific KG provide more relevant and useful knowledge grounding for \textit{CE-RAG4EM} than those from the commonsense KG, while \textbf{Example 3} shows that the triples retrieved from the commonsense KG provide more relevant and useful knowledge grounding for \textit{CE-RAG4EM} than those from the domain-specific KG. 

\begin{tcolorbox}[boxrule=0.8pt, colback=blue!5, breakable]
\noindent\textbf{Example 2: KG20C works \textit{better} than Wikidata}

    \textbf{EM Pair}: \textit{Entity 1}: Entity 1: title: towards self-tuning data placement in parallel database systems; authors: m lee , m kitsuregawa , b ooi , k tan , a mondal; venue: sigmod conference; year: 2000. \textit{Entity 2}: title: automating physical database design in a parallel database system; authors: j rao , c zhang , g lohman , n megiddo; venue: nan; year: 2002.0

    \textbf{Ground Truth}: 0 (No, Non-Matched)
    
    \textbf{Retrieved Wikidata Knowledge:} (Symposium on Principles of Database Systems, instance of, academic conference series); (Spatial Query Server, instance of, software)
    
    \textbf{Retrieved KG20C Knowledge:} (masaru kitsuregawa, author\_write\_paper, Towards self-tuning data placement in parallel database systems); (mong li lee, author\_write\_paper, Towards self-tuning data placement in parallel database systems)
    
    \textbf{CE-KG-RAG4EM with Wikidata Triples Output}: \textcolor{LimeGreen}{No}; \textit{The additional information provided consists of triples: (Symposium on Principles of Database Systems, instance of, academic conference series); (Spatial Query Server, instance of, software). Neither of these triples directly relates to the content of the two entities provided. They do not mention SIGMOD or any relevant conferences or topics that would assist in making a comparison between the two entities. Thus, we will not use the additional information in our decision-making.} Based on the analysis of titles, authors, venues, and years, it is clear that the two entities represent different works in the field of database systems. 
    
    \textbf{CE-KG-RAG4EM with KG20C Triples Output}: \textcolor{LimeGreen}{No}; \textit{These triples confirm the authorship of Entity 1 but do not provide any information about Entity 2. Since the authors of Entity 2 do not appear in the additional information related to Entity 1, it reinforces the notion that these are different entities.} Based on the analysis of titles, authors, venues, years, and the additional information, we conclude that the two entities do not refer to the same entity.
\end{tcolorbox}

\begin{tcolorbox}[boxrule=0.8pt, colback=blue!5, breakable]
    \noindent\textbf{Example 3: Wikidata works \textit{better} than KG20C}

    \textbf{EM Pair}: \textit{Entity 1}: title: a query language for multidimensional arrays : design , implementation , and optimization techniques; authors: l libkin , r machlin , l wong; venue: sigmod conference; year: 1996. \textit{Entity 2}: title: a query language for multidimensional arrays : design; authors: l libkin , r machlin , l wong; venue: implementation , and optimization techniques , sigmod conf. , \& hellip ; ,; year: nan

    \textbf{Ground Truth}: 1 (Yes, Matched)
    
    \textbf{Retrieved Wikidata Knowledge:} (MultiDimensional eXpressions, instance of, query language); (Perl Data Language, has use, science)
    
    \textbf{Retrieved KG20C Knowledge:} (Efficient organization of large multidimensional arrays, paper\_in\_domain, Database); (Efficient organization of large multidimensional arrays, paper\_in\_domain, Database theory)
    
    \textbf{CE-KG-RAG4EM with Wikidata Triples Output}: \textcolor{LimeGreen}{Yes}; \textit{The first triple is relevant as it confirms the existence of a query language for multidimensional data, which aligns with the subject matter of both entities. The second triple is less relevant to our specific entities since it pertains to Perl Data Language.} Despite the discrepancies in the title and venue of Entity 2, the consistent authorship and the partial overlap in the title and venue suggest that both entities refer to the same scholarly work, albeit with one being a more complete representation than the other.

    \textbf{CE-KG-RAG4EM with KG20C Triples Output}: \textcolor{red}{No}; The additional information provided consists of triples related to "Efficient organization of large multidimensional arrays." However,\textit{this information does not directly relate to the titles, authors, or publication venue of the two entities in question.} Therefore, it does not provide useful context or relevance for the decision-making process. Given the significant differences in the titles, the ambiguity in the venue for Entity 2, and the lack of a publication year, it is reasonable to conclude that Entity 1 and Entity 2 do not refer to the same entity.
\end{tcolorbox}

\subsection{Failure Analysis} \label{appsub:failure-analysis}

We characterize the main failure modes of \textit{CE-RAG4EM} based on patterns observed across datasets.

\noindent\textbf{E1: False positives from weakly relevant retrieved context.} On datasets with mixed numeric and textual attributes (e.g., \textsf{AMGO}, \textsf{WAAM}), vector search sometimes retrieves entities that are topically related but factually distinct from the queried records. 
As shown in failure example 1, when the LLM is presented with this weakly related context alongside ambiguous record pairs from \textsf{AMGO}, it may over-anchor to surface-level cues in the retrieved text and incorrectly predict a match.
In particular, the LLM ranked this retrieved contextual knowledge as lower relevance, which may not help in the decision-making, but \textit{CE-RAG4EM} with Top-2 ranked PID still outputs the incorrect answer in these specific cases.

\begin{tcolorbox}[boxrule=0.8pt, colback=blue!5, breakable]
\noindent\textbf{Failure Example 1: Weakly Relevant Knowledge}

    \textbf{EM Pair}: \textit{Entity 1}: title: punch ! master landscape \& home design; manufacturer: punch ! software; price: 99.99.  \textit{Entity 2}: title: punch software 90500 home design studio mac; manufacturer: punch software; price: 135.99

    \textbf{Ground Truth}: 0 (No, Not Matched)
    
    \textbf{Retrieved Contextual Knowledge:} P7514 (landscape architect: person or entity responsible for the landscape design of the item); P4435 (snap package: Snapcraft-based Linux software install)
    
    \textbf{LLM Output}: \textcolor{LimeGreen}{No}; While both entities share the same manufacturer and contain overlapping keywords related to home design, they differ in title focus, potential product type, and pricing. These differences suggest that they likely refer to different products rather than the same entity.
    
    \textbf{CE-KG-RAG4EM Output}: \textcolor{red}{Yes}; \textit{The additional information: P7514 (landscape architect) relates to design, where both products may serve similar purposes in design.} Given the shared manufacturer, the common purpose of design, and the contextual relevance of the additional information, it is reasonable to conclude that both entities refer to the same underlying product, albeit possibly different versions or editions.
\end{tcolorbox}

\noindent\textbf{E2: False negatives from absent or irrelevant retrieval.} On datasets with sparse or abbreviated attributes (e.g., \textsf{BEER}), the query constructed from the record pair may fail to retrieve a relevant Wikidata entity, leaving the LLM with no useful grounding. In these cases, \textit{CE-RAG4EM} effectively reduces to standard LLM-EM, and the recall improvement is limited.  This explains why \textsf{BEER} shows high precision but comparatively lower recall in Table~\ref{tab:cerag4em_llmem_metrics}.

\begin{tcolorbox}[boxrule=0.8pt, colback=blue!5, breakable]
\noindent\textbf{Failure Example 2: Absence of Relevant Knowledge}

    \textbf{EM Pair}: \textit{Entity 1}:  Beer\_Name: Jazz Beer Tarraco Amber; Brew\_Factory\_Name: Tarraco Beer; Style: American Amber / Red Ale; ABV: 6.90 \%.  \textit{Entity 2}: Beer\_Name: Jazz Beer Tarraco Amber; Brew\_Factory\_Name: Ca l\_Arenys - Cervezas Guineu; Style: Amber Ale; ABV: 6.90 \%

    \textbf{Ground Truth}: 1 (Yes, Matched)
    
    \textbf{Retrieved Contextual Knowledge:} (amber ale, subclass of, pale ale) $\rightarrow$ (amber ale, instance of, beer style); (American Beverage Association, instance of, company) $\rightarrow$ (American Beverage Association, instance of, trade association)
    
    \textbf{LLM Output}: \textcolor{red}{No}; The Beer\_Name is the same, and the ABV is identical, which are strong points for a match. However, the differing brew factory names and slightly different styles raise questions about whether they are indeed the same product. Since the brew factory names are different, this indicates they are likely produced by different entities.
    
    \textbf{CE-KG-RAG4EM Output}: \textcolor{red}{No}; The provided additional information related to beer styles and categories. Information about the American Beverage Association and its relation to the United States. \textit{The additional information does not provide direct relevance to the specific beers in question, particularly since it focuses on broader categories and historical context rather than specific entities.} Considering the strong discrepancy in the brewing factories and the potential difference in style classifications, despite the same beer name and ABV, it is reasonable to conclude that these are likely not the same entity.
\end{tcolorbox}

\end{document}